\documentclass[a4paper,12pt]{article}
\pdfoutput=1
\usepackage{graphicx, rotating,a4wide}
\usepackage{hyperref}
\usepackage{ifpdf,colortbl}
\usepackage{cite}

\ifx\pdfoutput\undefined
\usepackage[dvips,bookmarks=false]{hyperref}	
\else
\usepackage{hyperref}	
\fi
\hypersetup{colorlinks,bookmarksopen,bookmarksnumbered,citecolor=verdes,
linkcolor=blus,pdfstartview=FitH,urlcolor=rossos}
\def\hhref#1{\href{http://arxiv.org/abs/#1}{#1}} 

\def\lsim{\mathrel{\rlap{\lower3pt\hbox{\hskip0pt$\sim$}}
   \raise1pt\hbox{$<$}}}         
\def\gsim{\mathrel{\rlap{\lower4pt\hbox{\hskip1pt$\sim$}}
   \raise1pt\hbox{$>$}}}         

\definecolor{verdes}{cmyk}{0.92,0,0.59,0.4}

\def\hhref#1{\href{http://arxiv.org/abs/#1}{#1}} 

\def\m@th{\mathsurround=0pt }
\def\leftrightarrowfill{$\m@th \mathord\leftarrow \mkern-6mu
        \cleaders\hbox{$\mkern-2mu \mathord- \mkern-2mu$}\hfill
        \mkern-6mu \mathord\rightarrow$}

\def\overleftrightarrow#1{\vbox{\ialign{##\crcr
        \leftrightarrowfill\crcr\noalign{\kern-1pt\nointerlineskip}
        $\hfil\displaystyle{#1}\hfil$\crcr}}}

\makeatletter
\def\art{\@ifnextchar[{\eart}{\oart}}
\def\eart[#1]#2#3#4#5#6{{\rm #2}, {#3 #4} {\rm (#6) #5} [arXiv:{\hhref{#1}}]}
\def\hepart[#1]#2{{\rm #2, arXiv:\hhref{#1}}}
\newcommand{\oart}[5]{{\rm #1}, {#2 #3} {\rm (#5) #4}}

\makeatother

\definecolor{rosso}{cmyk}{0,1,1,0.4}
\definecolor{rossos}{cmyk}{0,1,1,0.55}
\definecolor{rossoc}{cmyk}{0,1,1,0.2}
\definecolor{blu}{cmyk}{1,1,0,0.3}
\definecolor{blus}{cmyk}{1,1,0,0.6}
\definecolor{bluc}{cmyk}{1,1,0,0.1}
\definecolor{verde}{cmyk}{0.92,0,0.59,0.25}
\definecolor{verdec}{cmyk}{0.92,0,0.59,0.15}
\definecolor{verdes}{cmyk}{0.92,0,0.59,0.4}
\definecolor{grigio}{cmyk}{0,0,0,0.07}
\definecolor{rosa}{cmyk}{0,0.1,0.1,0.02}
\definecolor{rosino}{cmyk}{0,0.05,0.05,0.02}
\definecolor{rosas}{cmyk}{0,0.3,0.25,0.05}
\definecolor{celeste}{cmyk}{0.1,0,0,0.02}
\definecolor{giallino}{cmyk}{0,0,0.4,0.02}
\definecolor{rosso}{cmyk}{0,1,1,0.4}
\definecolor{rossos}{cmyk}{0,1,1,0.55}
\definecolor{rossoc}{cmyk}{0,1,1,0.2}
\definecolor{blu}{cmyk}{1,1,0,0.3}
\definecolor{bluc}{cmyk}{1,1,0,0.1}
\definecolor{blucc}{cmyk}{0.7,0.5,0,0}
\definecolor{viola}{cmyk}{0,1,0,0.6}
\definecolor{viola2}{cmyk}{0,1,0.2,0.6}
\definecolor{verde}{cmyk}{0.92,0,0.59,0.25}
\definecolor{verdec}{cmyk}{0.92,0,0.59,0.15}
\definecolor{verdes}{cmyk}{0.92,0,0.59,0.4}
\definecolor{verdino}{cmyk}{0.12,0,0.09,0.05}
\definecolor{giallo}{cmyk}{0,0,1,0}
\definecolor{gialloverde}{cmyk}{0.44,0,0.74,0}

\font\tenrsfs=rsfs10 at 12pt
\font\sevenrsfs=rsfs7
\font\fiversfs=rsfs5
\newfam\rsfsfam
\textfont\rsfsfam=\tenrsfs
\scriptfont\rsfsfam=\sevenrsfs
\scriptscriptfont\rsfsfam=\fiversfs
\def\mathscr#1{{\fam\rsfsfam\relax#1}}

\newcommand{\gev}{\hbox{\rm\,GeV}}

\newcommand{\be}{\begin{equation}}
\newcommand{\ee}{\end{equation}}
\newcommand{\beq}{\begin{equation}}
\newcommand{\eeq}{\end{equation}}

\def\shat{\ifmmode \hat{s}\else $\hat{s}$\fi}
\def\gp2{{g'}^2}
\def\g2{g^2}
\def\g32{g_s^2}

\newcommand{\newc}{\newcommand}

\newc{\ie}{{\it i.e.}}
\newc{\etal}{{\it et al.}}
\newc{\mev}{\hbox{\rm\,MeV}}
\newc{\tev}{\hbox{\rm\,TeV}}
\newc{\xpb}{\hbox{\rm\, pb}}
\newc{\xfb}{\hbox{\rm\, fb}}

\newc{\G}{{\cal G}}
\newc{\h}{{\cal H}}
\newc{\D}{{\cal D}}
\newc{\E}{{\cal E}}
\newc{\cB}{{\rm BR}}
\newc{\mtop}{M_t}
\newc{\mbot}{m_b}
\newc{\mz}{M_Z}
\newc{\mw}{M_W}
\newc{\alphasmz}{\alpha_s(M_Z)}
\newc{\swsq}{\sin^2\theta_W}
\newc{\cwsq}{\cos^2\theta_W}
\newc{\tw}{\tan\theta_W}
\newc{\cw}{\cos\theta_W}
\newc{\sw}{\sin\theta_W}
\newc{\BR}{\hbox{\rm BR}}
\newc{\zbb}{Z\to b\bar}
\newc{\Gb}{\Gamma (Z\to b\bar b)}
\newc{\Gh}{\Gamma (Z\to \hbox{\rm hadrons})}
\newc{\sgn}{\mbox{sgn}}

\def\eq#1{eq.~(\ref{#1})}
\def\fig#1{fig.~\ref{#1}}

\newcounter{mysubequation}[equation]

\newcommand{\TeV}{\,\mathrm{TeV}}
\newcommand{\GeV}{\,\mathrm{GeV}}

\def\beq{\begin{equation}}
\def\eeq{\end{equation}}
\def\bea{\begin{eqnarray}}
\def\eea{\end{eqnarray}}
\def\slashchar#1{\setbox0=\hbox{$#1$}           
   \dimen0=\wd0                                 
   \setbox1=\hbox{/} \dimen1=\wd1               
   \ifdim\dimen0>\dimen1                        
      \rlap{\hbox to \dimen0{\hfil/\hfil}}      
      #1                                        
   \else                                        
      \rlap{\hbox to \dimen1{\hfil$#1$\hfil}}   
      /                                         
   \fi}                                         %
\catcode`@=11
\long\def\@caption#1[#2]#3{\par\addcontentsline{\csname
  ext@#1\endcsname}{#1}{\protect\numberline{\csname
  the#1\endcsname}{\ignorespaces #2}}\begingroup
    \small
    \@parboxrestore
    \@makecaption{\csname fnum@#1\endcsname}{\ignorespaces #3}\par
  \endgroup}
\catcode`@=12

\newcommand{\mstop}{m_{S}}
\newcommand{\mstopl}{m_{{\tilde t}_1}}
\newcommand{\mstoph}{m_{{\tilde t}_2}}
\newcommand{\mstopR}{m_{{\tilde t}_R}}
\newcommand{\mstopL}{m_{{\tilde t}_L}}
\newcommand{\mhu}{m_{H_u}}

\begin{document}

\baselineskip=18pt

\begin{titlepage}
\begin{flushright}
\end{flushright}
\vspace{.3in}

\begin{center}
{\LARGE\color{black}\bf The light stop window}
\vspace{0.5cm}\\
{\bf Antonio Delgado$^{a}$,
Gian F. Giudice$^{b}$,  
Gino Isidori$^{b,c}$, \\
Maurizio Pierini$^{b}$, 
Alessandro Strumia$^{d,e}$}\\
\vspace{0.5cm}
{\it $(a)$ Department of Physics, University of Notre Dame, Notre Dame IN 46556, USA}\\
{\em $(b)$ {CERN, Theory Division, CH--1211 Geneva 23,  Switzerland}}\\
{\em $(c)$ {INFN, Laboratori Nazionali di Frascati, I-00044 Frascati, Italy}}\\
{\it $(d)$ Dipartimento di Fisica, Universit{\`a} di Pisa and INFN Sez. Pisa, Pisa, Italy}\\
{\it  $(e)$ National Institute of Chemical Physics and Biophysics, Tallinn, Estonia}\\
\vspace{0.5cm}

\end{center}
\vspace{.8cm}

\centerline{\large\bf Abstract}
\begin{quote}\large
We show that a right-handed stop in the 200--400 GeV mass range, together with a nearly degenerate neutralino and, possibly, a gluino below 1.5 TeV, follows from reasonable assumptions, is consistent with present data, and offers interesting discovery prospects at the LHC. Triggering on an extra jet produced in association with stops allows the experimental search for stops even when their mass difference with neutralinos is very small and the decay products are too soft for direct observation. Using a razor analysis, we are able to set stop bounds that are stronger than those published by ATLAS and CMS.
\end{quote}

\bigskip
\bigskip

\end{titlepage}

\section{Introduction}

Supersymmetry has been significantly cornered by LHC searches. The discovery of the Higgs boson at 126~GeV~\cite{HiggsATLAS,HiggsCMS}
 and the direct limits on sparticles rule out most of the natural implementations of low-energy supersymmetry, at least in their simplest versions~\cite{nat}. Pockets of parameter space still survive, but their exploration requires the 14-TeV phase of the LHC. At this stage, it is appropriate to examine the available experimental data and look for hints that can guide us towards special regions where supersymmetry may still hide. 

In this paper, we point out that there is a window of supersymmetric parameters that {\it (i)} are well consistent with all collider data and flavor constraints, {\it (ii)} naturally emerge from RG evolution of simple UV completions, {\it (iii)} predict the correct thermal abundance for dark matter (DM), and {\it (iv)} give observable signals at LHC14. In this special window, the supersymmetric mass spectrum has the following properties:
\begin{itemize}
\item
The lightest stop is mostly right-handed and its mass is in the range $\mstopl = 200$--$400\gev$.
\item
The heavy, mostly left-handed, stop has a much larger mass (in the 1--2 TeV range), but it is correlated with the light stop in such a way that their geometric average is
$\mstop \equiv (\mstoph \mstopl)^{1/2}  \approx 500$--$600\gev$.
\item
The stop trilinear term is large, such that\footnote{This configuration is  known as the ``maximal mixing" case, 
 although it does not necessarily imply a large mixing between the two stop mass eigenstates, as discussed in sect.~\ref{sec1}.} $A_t^2 \approx 6 \mstop^2$. 
 \item
 The gluino mass is below about 1.5~TeV.
 \item
 The lightest neutralino has a mass slightly smaller than the lightest stop, by an amount of about 30--$40\gev$.
 \end{itemize}
 
 In section~\ref{sec1} we will give several arguments that lead to the mass spectrum described above. None of them is sufficiently compelling to select conclusively the sparticle masses but these arguments, taken together, give circumstantial evidence in favor of our choice of parameters. Our conclusion is based on the following considerations:
\begin{itemize}
\item
We choose the values of the stop parameters that minimize the average stop mass, while leading to a Higgs mass of about 126~GeV. The resulting stop mass spectrum, although not strictly natural, has the advantage of reducing the amount of unnaturalness forced upon supersymmetry by present LHC data. 
\item
Stops affect the rates for $gg\to h$ and $h\to \gamma \gamma$. Experimental data are at present not sufficiently constraining, but will soon play an important role in selecting stop parameters. In particular, we find that stop contributions to $gg\to h$ and $h\to \gamma \gamma$ exactly cancel for $\mstoph \approx 6 \mstopl$, under the conditions preferred by the Higgs mass value.
\item
Flavor constraints rule out a light left-handed stop, but are consistent with a light right-handed stop. We also show that, if the CKM matrix is the only source of flavor violation and the higgsino is relatively light, supersymmetric contributions can improve the agreement with the measurements of $\epsilon_K$, while being compatible with $B \to X_s \gamma$. This is achieved for a large mass splitting between left-handed and right-handed stops.
\item
A large splitting between left-handed and right-handed stop masses naturally emerge from RG evolution, as long as the gluino is not too heavy. Moreover, once we require a large stop mixing parameter at low energy, we find an upper bound on the gluino mass.
\item
A light stop can be very helpful to obtain the right dark-matter relic density, which is typically too large for B-ino LSP or too small for higgsino or W-ino LSP in generic supersymmetric models. The process of  coannihilation selects the preferred value of the mass difference between stop and neutralino.  
\end{itemize}

After we have determined the favorable region for sparticle masses, we study in sect.~\ref{sec3} the experimental strategies for discovery at the LHC. The phenomenology of supersymmetric frameworks with light stop has been discussed at length in 
the recent literature (see e.g.~ref.~\cite{Barbieri:2009ev,Papucci:2011wy,Brust:2011tb,Han:2012fw,Espinosa:2012in,Carena:2008rt}). 
The search of light stops has also been the 
focus of recent dedicated experimental analyses by both ATLAS and CMS~\cite{Exppages}. However, 
the peculiarity of our scenario is the near mass-degeneracy between stop and neutralino, which makes experimental identification especially arduous (see ref.~\cite{Chou:1999zb,Hiller:2009ii,Alves:2012ft,Kilic:2012kw} for previous attempts to address this problem).

On the one hand, we show that the decay $\tilde t \to Nb\ell \nu_\ell$, which has been neglected in the present exclusive experimental searches, can dominate over the more traditional decay $\tilde t \to c N$, especially if the mass difference between $\tilde t$ and $N$ is not too small. The four-body decay process has the advantage of producing observable leptons in the final state, leading 
to a possibly higher signal/background ratio in exclusive searches. 
On the other hand, by an explicit simulation of this decay channel and the analysis of presently available data, 
we show that the inclusive searches, and particularly the  CMS razor analysis,  already provides significant  constraints 
on this framework.

\begin{figure}[t]
\begin{center}
$$\includegraphics[width=0.45\textwidth]{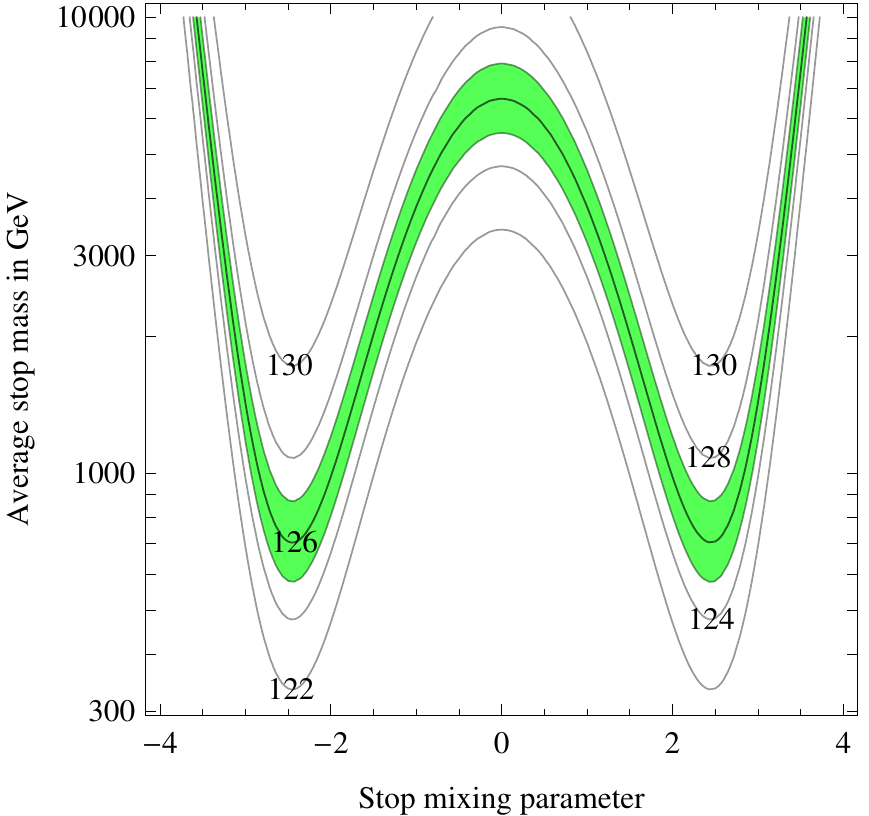}\qquad
\includegraphics[width=0.45\textwidth]{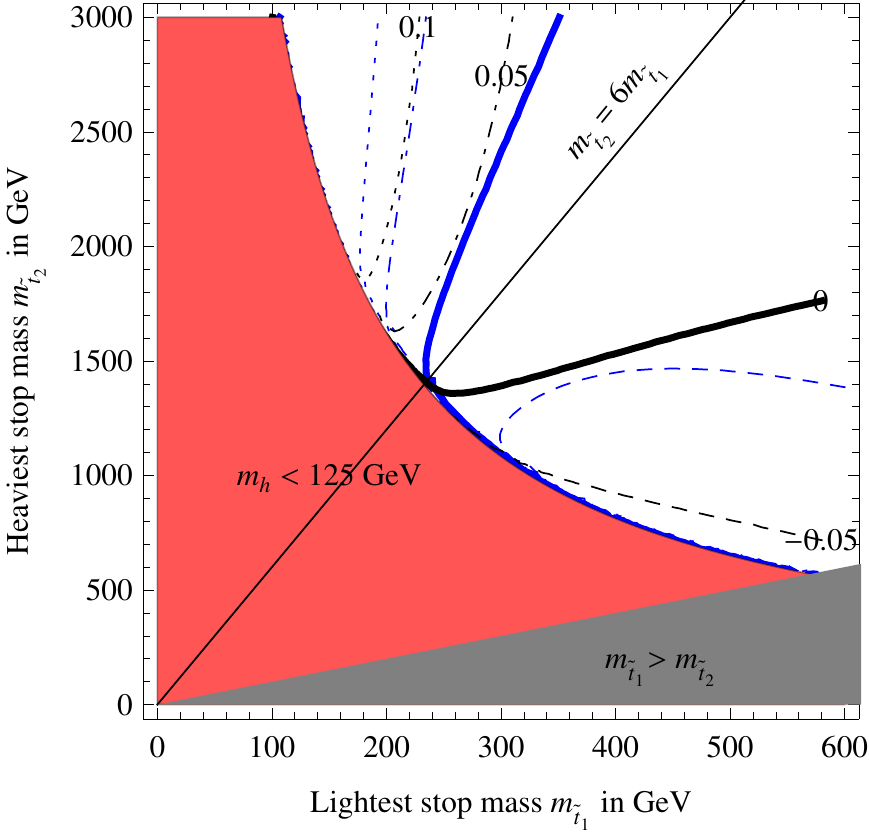}$$
\end{center}
\parbox{0.48\textwidth}{\caption{\label{fig:stop}\em
The Higgs mass in low-energy supersymmetry for large $\tan\beta\approx 20$.
The shaded region in the  $ (X_t,  \mstop)$ plane corresponds to the observed value of $m_h$. Higher-order corrections and the uncertainty in the top mass amount to an error of a few $\GeV$ in $m_h$.
}}\hspace{0.04\textwidth}
\parbox{0.48\textwidth}{\caption{\label{fig:stopA}\em 
The white region is the range in the $(\mstopl,\mstoph)$ plane allowed by the $m_h$ constraint, while shaded regions are excluded.
The full, dashed, and dotted lines  correspond to fixed values of $\Delta_t$, satisfying the $m_h$ constraint
with $|X_t| > \sqrt{6}$  (blue) or $|X_t| < \sqrt{6}$  (black).
}}
\end{figure}

\section{The light-stop window}\label{sec1}
\subsection{Constraints from the Higgs mass and decay rates}
The leading part of the supersymmetric prediction for the mass of the lightest Higgs boson  is 
\be
 m_h^2 = m^2_Z  \cos^2 2\beta +  \frac{3 y_t^2 m_t^2 }{4\pi^2} \left[ \log\left( \frac{\mstop^2}{m_t^2}\right) + 
X_t^2 \left( 1 - \frac{X^2_t}{12 }  \right) \right]+\cdots
\label{higgsmass}
\ee
where $X_t = (A_t +\mu\cot\beta)/\mstop$,  
$\mstop^2=\mstopl\mstoph$ is the average stop mass,  $y_t= m_t/v$ is the top-quark Yukawa coupling, 
and $v\approx 174$~GeV is the Higgs vev.
In fig.~\ref{fig:stop} we show the region of the $(X_t, \mstop)$ plane compatible with the observed Higgs mass
(for $\tan\beta\gg1$), 
including also the leading two-loop corrections to the Higgs mass not shown in \eq{higgsmass}.
The lightest average stop mass that can lead to the observed Higgs mass
is obtained for
\be
 \mstop \approx  500~{\rm GeV}  \qquad  {\rm and } \qquad   X^2_t   \approx 6~.
 \label{eq:minstop}
 \ee
 We focus on  such configuration, the so-called ``maximal mixing" case, since it reduces the fine-tuning in electroweak symmetry breaking and can lead to observable signals.

Further constraints on $X_t$ and the stop masses can be obtained by examining the corrections 
to the $h\to\gamma\gamma$ and $h\leftrightarrow  gg$ rates:
\be
\frac{\Gamma(h \leftrightarrow  gg)}{\Gamma (h\leftrightarrow  gg)_{\rm SM} }=  (1+\Delta_t)^2~, \qquad 
\frac{\Gamma(h \to \gamma \gamma)}{\Gamma (h\to \gamma \gamma)_{\rm SM} } =  (1-0.28 \Delta_t)^2~,
\ee 
where, in the limit in which we decouple the pseudoscalar Higgs, we find  
\be
\Delta_t \approx \frac{m_t^2}{4} \left( \frac{1}{\mstopl^2} +  \frac{1}{\mstoph^2} -  \frac{X_t^2}{\mstop^2} \right) .
\ee
Present data (fitted in the context of the SM plus light stops) give~\cite{fit}
\be
\Delta_t = -0.04\pm0.11
\ee
and do not yet imply a significant constraint, as it is clear from \fig{fig:stopA} where we 
plot iso-curves of $\Delta_t$  after imposing the $m_h$ requirement. The situation will improve in the future. Note that no deviations from the 
SM ($\Delta_t \approx 0$) are obtained for $\mstoph\approx 6\mstopl$
if we insist on having $X_t^2 \approx 6$.
%

 A few comments are in order: 
\begin{itemize}
\item{}
An independent indication of a large splitting between $\mstoph$ and $\mstopl$ can be obtained if we 
assume that $A_t$ is not significantly larger than the trace of the stop mass matrix.
Assuming $A^2_t < a (\mstopl^2+\mstoph^2)$, then  (for large $\tan\beta$) $X^2_t$  is bounded by 
\be
X_t^2  < a \frac{  \mstopl^2+\mstoph^2 }{ \mstopl \mstoph }  
~ \stackrel{ r \ll 1 }{\simeq} ~  \frac{ a }{r}~, \quad \qquad r = \frac \mstopl \mstoph ~.
\ee
Vacuum stability arguments imply $a < 3$ (assuming $m^2_{H_u}  \ll \mstoph^2$), but this does not allow us 
to deduce a significant constraint on $r$. However,  if $a\lsim1 $ 
(as naturally expected from RG arguments, see next section) then we are forced to assume small values of $r$
in order to reach  $X^2_t   \approx 6$.
\item{} Despite the large value of $X_t$, 
the mixing of the two stop eigenstates is  suppressed in the limit $r\ll 1$:
\be
\theta_t = \frac{1}{2} \arcsin\left( \frac{2m_t \mstop X_t }{ \mstoph^2 - \mstopl^2} \right) ~
\stackrel{ r \ll 1 }{\simeq} ~  \frac{r X_t m_t}{\mstop} ~.
\label{eq:thetat}
\ee
So, in this limit, we can approximately identify the two mass eigenstates with the electroweak 
eigenstates. As we will show in the next section, it is natural to identify the lightest state
with an almost right-handed stop. 
Note also that  for $r \ll1$ the lightest stop mass is significantly lighter than the average stop mass in \eq{eq:minstop}: $r \approx 1/6$ corresponds to
 $\mstopl \approx 200$~GeV.
 \end{itemize}

\begin{figure}[t]
\begin{center}
$$\includegraphics[width=0.46\textwidth,height=0.42\textwidth]{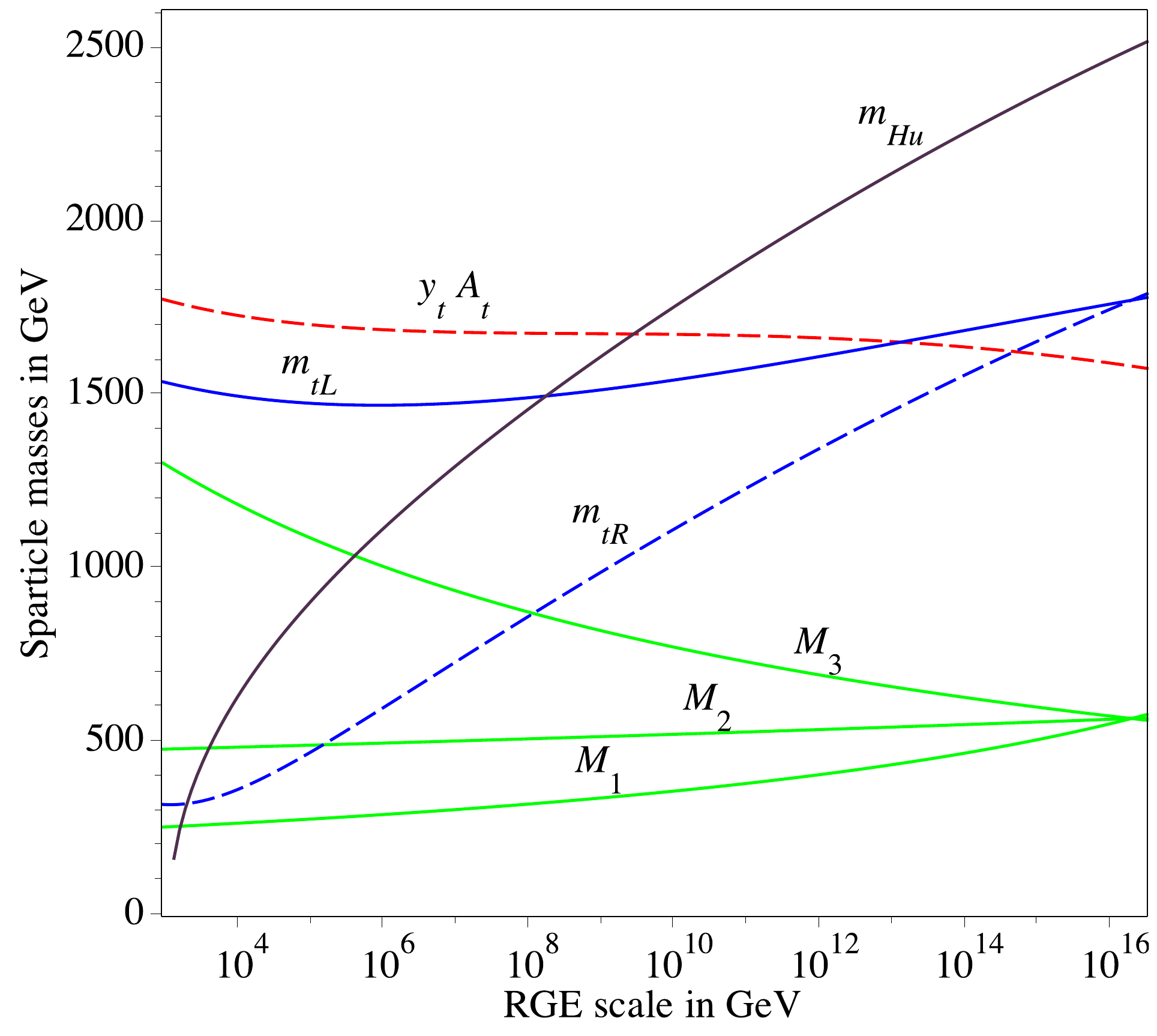}\qquad
\includegraphics[width=0.46\textwidth,height=0.44\textwidth]{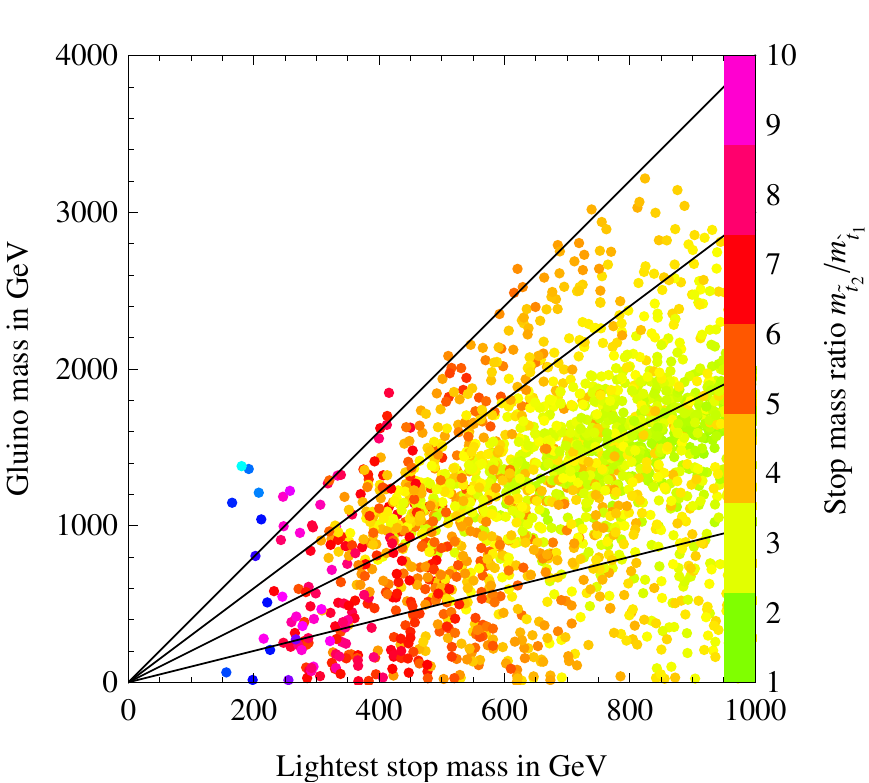}
$$
\end{center}
\parbox{0.48\textwidth}{\caption{\label{fig:RGE}
\em Illustrative example of
renormalization group evolution from the unification scale to the weak scale
of  gaugino masses $M_1$, $M_2$, $M_3$
(green curves), 
of the stop mass parameters
$\mstopL$  and $\mstopR$ (full and dashed blue curves, respectively), $y_t A_t$ (red dashed curve), $\mhu$ (black curve),
in a  configuration  leading to $\mstopR \ll \mstopL$ at the weak scale.  All masses
are in {\rm GeV} units and we assumed the MSSM.
}}\hspace{0.04\textwidth}
\parbox{0.48\textwidth}{\caption{\label{fig:gluino}
\em Gluino and light-stop masses resulting from 
a scan of the parameter space assuming 
universal scalar and gaugino masses, and  the 
condition $|A_t| < 3 m_0$, at the GUT scale. 
All points satisfy the $m_h\approx 126~{\rm GeV}$ constraint and are colored according to the value of
$\mstoph/\mstopl$, as indicated on the right-handed axis. For illustrative purposes lines  
corresponding to  $M_3/\mstopl=1,2,3,4$ are also shown.
}}
\end{figure}

\subsection{Constraints from the RG evolution}

A numerically large splitting between $\mstopL$ and $\mstopR$ naturally arises
from the evolution under renormalization-group equations (RGE), provided scalar masses are significantly 
larger than gaugino masses at the high scale~\cite{ASold}. This can be understood by looking 
at the one-loop RGE for third generation squark masses  and $\mhu$.
Neglecting off-diagonal flavor-mixing terms we have
\bea
8\pi^2~\frac{d \mstopL^2}{d\log \mu} &=&  y_t^2 Y_t -\frac{16}{3} g_3^2  M_3^2 - 3 g_2 M_2^2 -\frac{1}{15} g_1^2 M_1^2 ~, \\
8\pi^2~\frac{d \mstopR^2}{d\log \mu} &=&  2\, y_t^2 Y_t -\frac{16}{3} g_3^2  M_3^2 - \frac{16}{15}  g_1^2 M_1^2 ~, \\
8\pi^2~\frac{d \mhu^2}{d\log \mu} &=& 3\, y_t^2 Y_t - 3 g_2 M_2^2 -\frac{3}{5} g_1^2 M_1^2 ~, 
\eea
where $\mu$ is the renormalization scale, and 
\be
Y_t = \mstopL^2 +\mstopR^2 + \mhu^2 + A_t^2~.
\ee
The RG evolution of the stop masses depends mainly on two effects: the QCD term
 ($g_3^2  M_3^2$) and the Yukawa term ($y_t^2 Y_t$). If  we take   $M_3$, $\mstopL$ and $A_t$ to be comparable and in the range 1--2
 TeV  at the weak scale (in order to fulfill the $m_h$ constraints and 
the experimental bounds on the gluino mass), we find that: i) QCD and Yukawa terms compensate to a 
large extent in the running of $\mstopL$; ii) the Yukawa term is  dominant during most 
of the running of $\mstopR$, leading to $\mstopR \ll \mstopL$ at the weak scale starting 
from the initial condition $\mstopR = \mstopL$ at some high scale; iii) the Yukawa term is always 
dominant in the running of $\mhu^2$, which naturally becomes negative at the weak scale.

An illustrative spectrum is shown in fig.~\ref{fig:RGE}a, where we required $\mhu^2 = -m_Z^2/2$,
$M_3=1.3$~TeV, and $\mstopR < 300$~GeV
at the weak scale, and adjusted $A_t$ in order to achieve the condition $\mstopR = \mstopL$ at  
$2\times 10^{16}$~GeV. The corresponding weak-scale configuration is  consistent with all the existing experimental bounds, 
with the condition $m_h = 126\gev$ (assuming $\tan\beta \gsim 5$), and with a light stop below 300~GeV.
A few percent tuning in the initial values of $\mhu^2$ and the higgsino mass $\mu$ is necessary in order to achieve the 
correct pattern of electroweak symmetry breaking, but this is unavoidable in the minimal supersymmetric model with 
$m_h= 126\gev$. The soft-breaking terms  needed to reach this low-energy configuration 
require an initial splitting $m^2_{\rm squarks} /m^2_{\rm gauginos}  \sim 10$ at the high-energy scale. All three generations of squarks can be degenerate at the high scale, since the separation of the right-handed stop is fully driven by the dynamics of the low-energy degrees of freedom. In this case the squarks of the first two generations would have a mild RGE evolution
(reaching low-energy values slightly above 2 TeV for the illustrative configuration shown in fig.~\ref{fig:RGE}a).

The dynamical separation between $\mstopR$ and $\mstopL$, together with the generation of a large $X_t$, from high-scale RG running naturally occurs only in a limited range of gluino masses. This can be understood by inspecting the expressions of $A_t$, 
$\mstopR$ and $\mstopL$  at the weak scale in models with a universal scalar mass $m_0$ and trilinear coupling $A_0$ at the GUT scale,
\bea
 A_t &\approx& 0.3 A_0 + 0.8 M_3 ~,  \\
m_{\tilde{t}_R}^2  &\approx&  0.5 M_3^2 - 0.07 A_0^2 - 0.10 A_0 M_3 + 0.3 m_0^2 ~, \\
m_{\tilde{t}_L}^2  &\approx&  0.7 M_3^2 - 0.03 A_0^2 - 0.05 A_0 M_3 + 0.7 m_0^2 ~,
\label{bb}
\eea
where $M_3$ is the gluino mass at the weak scale. From these equations it is clear that if $M_3 \ll  m_0, |A_0|$ a large splitting among the two stop masses and a large $X_t$ are obtained only for unnaturally large values of  $|A_0|/ m_0$.
Similarly, maximal mixing and large splitting cannot be obtained if $M_3 \gg  m_0, |A_0|$. 
The upper bound on $M_3$, which is particularly important for the LHC searches, is quantified in fig.~\ref{fig:gluino}, 
where we show the points satisfying the $m_h$ constraint in the $\mstopl$--$M_3$ plane.  
As can be seen, the gluino must satisfy the approximate upper bound $M_3 \lsim 4 \mstopl$, that implies 
 $M_3 \lsim 1.6$~TeV for the range of $\mstopl$  ($\mstopl \lsim 400$~GeV) corresponding to a large $\mstoph/\mstopl$ ratio.

As anticipated, in this framework $\mhu^2$ naturally becomes negative at the weak scale and the 
$\mu$ term must be properly adjusted to reproduce the correct value of $m_Z$. Assuming 
universal scalar masses at the high scale, $\mhu^2$ runs very negative at the week scale, 
implying $|\mu| >  \mstopR$, or heavy higgsinos. Alternatively, we can consider  
a scenario as in  fig.~\ref{fig:RGE}, with non-universal boundary conditions, where $|\mu| < \mstopR$ and thus  
higgsinos are lighter than the right-handed stop. The two cases lead  to a  rather different
phenomenology for flavor, dark matter, and LHC searches.

 \subsection{Constraints from flavor physics}
If some of the gauginos or higgsinos are not too heavy, 
a light stop can have a significant impact on low-energy flavor-physics observables.
On general grounds, even if gauginos and higgsinos are in the several-TeV domain, sizable misalignments 
in flavor space between quark and squark mass matrices are excluded. 
Therefore we assume that the light stop is mostly right-handed and aligned in flavor space with the top quark. The remaining flavor violation is described by the usual CKM angles in charged currents.
An interesting and largely model-independent correlation (controlled only by the size of $|\mu|$
and the stop mass parameters) emerges between  
$\cB(B\to X_s \gamma)$ and $\epsilon_K$.

\begin{figure}[t]
\begin{center}
$$ \includegraphics[width=0.80\textwidth]{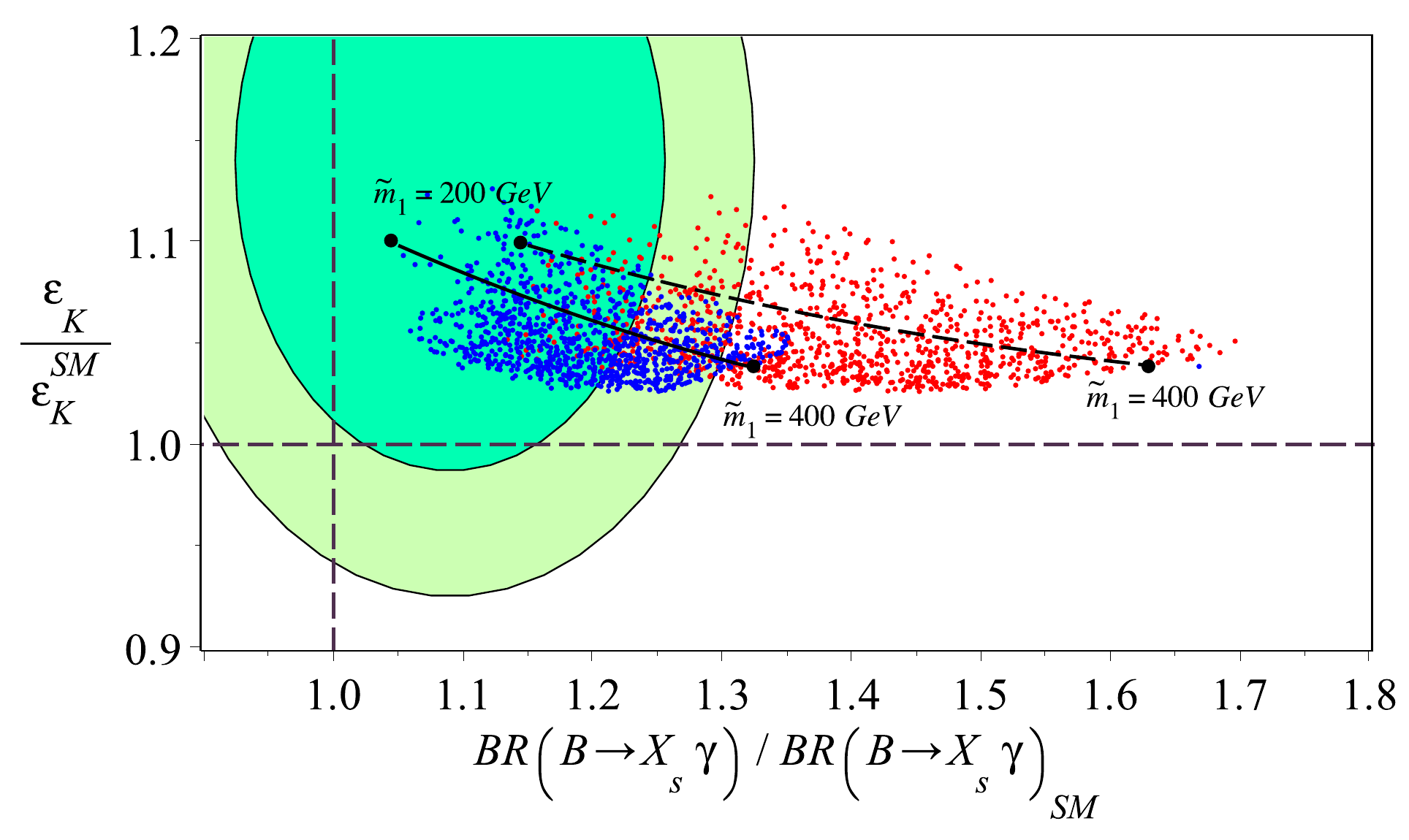}$$
\end{center}
\caption{\label{fig:flavor}\em
Correlation between $\cB(B\to X_s \gamma)$ and $\epsilon_K$.  
The two ellipses denote the $68\%$ and $90\%$ CL experimental range.
All points reproduce the observed Higgs mass. 
The two black curves are obtained 
varying $\mstopR$ between $200\GeV$ and $400\GeV$ (from left to right)
for $\mstop=500\GeV$, $\mu = 250\GeV$, and $\tan\beta=20$ (dashed curve) or 
$\tan\beta = 10$ (full curve).
The points are obtained varying the parameters in the range
$\mu =  [150-400]\GeV$ and $\mstopR=[200-400]\GeV$,  with $\mstop < 700 \GeV$ and 
$\tan\beta=20$ (red) or $\tan\beta = 10$  (blue). 
}
\end{figure}

In the limit in which we retain only the effect of higgsinos and flavor-aligned right-handed stop,
the deviations from the SM in these two observables are described by
\bea
\frac{\cB(B\to X_s \gamma)}{ \cB(B\to X_s \gamma)_{\rm SM} }  =
 1 - 2.5\, \Delta C_7 -0.7\,  \Delta C_8~, \\
\frac{\epsilon_K}{\epsilon_K^{\rm SM} } 
 =  1 + 1.9\, \frac{m_t^2} { \mstopR^2  } F_2\left(\frac{\mu^2}{\mstopR^2}\right)~,
 \eea
 where
 \be
\Delta C_{7,8} = \sin\theta_t \tan\beta \frac{  \mu m_t}{\mstopR^2} F_{7,8}^{LR} \left(\frac{\mu^2}{\mstopR^2}\right)
+   \frac{ m^2_t}{\mstopR^2}  F_{7,8}^{\rm RR} \left(\frac{\mu^2}{\mstopR^2}\right)~,
\label{eq:deltaC7}
\ee
and the normalization of the various loop functions 
(see Ref.~\cite{Gabrielli:1994ff,Ciuchini:1998xy} for the explicit expressions) is 
\be
 F_{7}^{\rm LR}(1) =  -\frac{2}{9}~, \quad F_{8}^{\rm LR}(1) = -\frac{1}{12}~, \quad 
F_{7}^{\rm RR}(1) =  \frac{5}{144}~, \quad F_{8}^{\rm RR}(1) = \frac{1}{48}~, \quad
F_2 (1) =   \frac{1}{12}~.
\ee 

For $B\to X_s \gamma$ we have expanded the result to first order in the stop-mixing angle.
Note that, even for $|\sin\theta_t|  \ll 1$, the first term in \eq{eq:deltaC7} can be sizable and can dominate over the second one, 
both because of the large value of the loop function $F_{7}^{\rm LR}$ and because the $m_h$ constraint favors $\tan \beta \gg 1$. 
As a result, the experimental constraint on $\cB(B\to X_s \gamma)$ puts a very stringent bound on the maximal value of 
$|\theta_t|$ for higgsino masses of $O(\mstopR$), providing a further argument in favor of a sizable hierarchy between the two stop mass eigenstates [see \eq{eq:thetat}]. The sign of the correction can be positive or negative,
depending on the relative sign of $\mu$ and $A_t$. The experimental data
favor a constructing interference with the SM 
amplitude: ${\rm BR}(B\to X_s \gamma)_{\rm exp}/{\rm BR}(B\to X_s \gamma)_{\rm SM} = 1.09\pm 0.11$.\footnote{This ratio is evaluated 
using the SM estimate from Ref.~\cite{Misiak:2006zs}, and a naive average of the HFAG result and the latest 
Babar result~\cite{Babar} on ${\rm BR}(B\to X_s \gamma)_{\rm exp}$. }

In the case of $\epsilon_K$, the correction is always positive and, in first approximation, is independent from the mixing angle. 
As a result, the present experimental constraint $\epsilon^{\rm exp}_K/\epsilon_K^{\rm SM} = 1.14 \pm 0.10$~\cite{UTfit} can be better satisfied if  $\mu$ is not too heavy. 

The correlation between the two observables is shown in fig.~\ref{fig:flavor}, where we restrict the attention to the value of ${\rm sgn}(\mu A_t)$  favored by $\cB(B\to X_s \gamma)$.
 As can be seen,  after imposing the $m_h$ constraint and  requiring  $|\mu| \lsim 400$~GeV,
 present data favor the configuration with $\mstopR \ll \mstopL$ that maximizes the correction to $\epsilon_K$ and
 minimizes the impact in   $\cB(B\to X_s \gamma)$.

 \subsection{Constraints from dark matter}

\begin{figure}[t]
\begin{center}
$$ \includegraphics[width=0.45\textwidth]{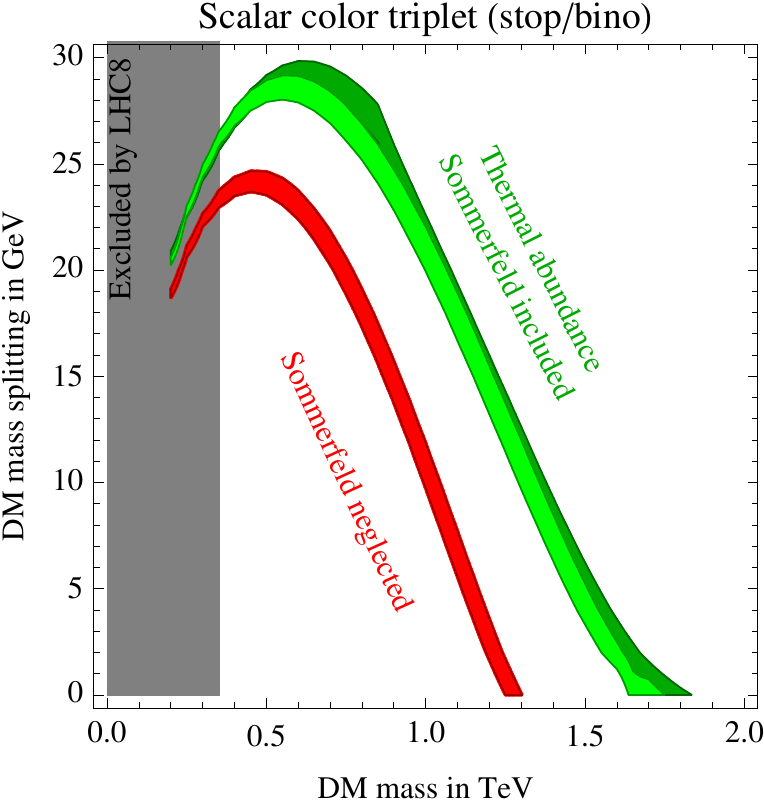}$$
\end{center}
\caption{\label{fig:dark}\em
Points in the supersymmetric parameter space that lead to the correct DM abundance. 
}
\end{figure}

A light stop offers the opportunity of curing the excessive relic abundance of B-ino LSP, generally encountered in supersymmetric models. Indeed, the DM cosmological abundance can be reproduced with a B-ino thermal relic that co-annihilates with stops if 
\beq m_{\tilde{t}_1}=  M_{\rm DM}   + \Delta M \qquad\hbox{with}\qquad \Delta M \approx 30 \GeV \label{eq:stopcoha}.\eeq
The relatively small mass difference  
arises  imposing that the average annihilation cross section equals
\beq 
\sigma v_{\rm cosmo}\equiv 
 (2.3\pm 0.1)\times10^{-26}\, {\rm cm}^3{\rm s}^{-1}\label{eq:stopDM}\eeq
 at the freeze-out temperature $T_f \approx   M_{\rm DM} /25$.
The dominant annihilation process is  $s$-wave stop annihilation into gluons (annihilation into quarks is $p$-wave suppressed):
\beq  \sigma (\tilde{t}_1 \tilde{t}_1^* \to gg)v = \frac{7\, g_3^4}{432\pi \, m_{\tilde{t}_1}^2}.\eeq
Averaging over the components of the DM system
($\tilde{t}$ and $\tilde{t}^*$ have 3 colors each, and the neutralino has 2 polarisations)
we get
\beq \sigma v_{\rm cosmo}= \sigma  (\tilde{t}_1 \tilde{t}_1^* \to gg)v\times \bigg[1 + \frac{e^{\Delta M/T}}{3(1+\Delta M/M)^{3/2} }\bigg]^{-1} .
\label{dmapp}
\eeq
The region where the DM abundance reproduces the cosmological values within
3 standard deviations is shown as a red band in \fig{fig:dark} (see also~\cite{CMSSMscan}).
In the figure we also show, as green band, the result of a more precise
computations that taks into account strong Sommerfeld corrections~\cite{Som}.
%
%
%
 
\section{Experimental signals} \label{sec3}
\subsection{Stop decay rates}

Beside gluino production, the most characteristic signal comes from the (mostly right-handed) light stop. Dark matter considerations motivate the searches for a stop that is near degenerate with the neutralino LSP, with a mass difference $\Delta M \equiv \mstopl - M_{\rm DM} \approx 30\GeV$. In this configuration, the stop is usually assumed to decay according to $\tilde{t}_1 \to c N$. Here we point out that four-body stop decays (not suppressed by flavor-changing neutral currents) can easily become competitive with the two-body flavor-violating decay. In the limit of small $\Delta M$, the relevant stop decay widths are
\be
  \Gamma(\tilde{t}_1 \to c N) = 
\frac{2g^2 \tan^2\theta_W \, \theta_{tc}^2 \, \Delta M^2}{9 \pi \, \mstopl} 
= 100~{\rm cm}^{-1}
\left( \frac{\theta_{tc}}{10^{-5}} \right)^2 \left(\frac{\Delta M}{30\GeV}\right)^2 
\left(\frac{400\GeV}{\mstopl}\right)~, 
\label{eq:2body} 
\ee
\be
\Gamma(\tilde{t}_1 \to N b\ell^+\nu_\ell) =
\frac{3\, g^6\tan^2\theta_W\, \Delta M^8}{70(6\pi)^5 M_W^4 \, m_t^2\, \mstopl }
=
28~{\rm cm}^{-1}
\left(\frac{\Delta M}{30\GeV}\right)^8 
\left(\frac{400\GeV}{\mstopl}\right)~,
\label{eq:4body} 
\ee
as well as
\be 
\Gamma(\tilde{t}_1 \to N b u \bar d) \approx
\Gamma(\tilde{t}_1 \to N b c \bar s) \approx
3\Gamma(\tilde{t}_1 \to N b\ell^+\nu_\ell) \qquad \ell =e,~\mu,~\tau~.
\label{eq:4body2} 
\ee

For the decay $\tilde{t}_1 \to c N$, the parameter  $\theta_{tc}$ is the effective stop--scharm mixing angle. 
In general, $\theta_{tc}$ is a free unknown parameter, since it depends on the flavor structure of the soft terms.  
Assuming that it vanishes at some high scale $\Lambda_{\rm UV}$, a non zero value is generated by RGE effects
due to the SM Yukawa couplings (even in absence of other sources of flavor violation)~\cite{Hiller:2008wp}.
In our scenario, where $\tilde{t}_1\approx \tilde{t}_R$, the leading effect comes from an induced $\tilde{t}_R$--$\tilde{c}_L$ mixing, which can be estimated as
\be
\theta^{\rm MFV}_{tc} \sim  \frac{y_t y_b^2 V_{cb}V_{tb}^* }{16\pi^2} \frac{v A}{ \tilde m^2}  \log\frac{\Lambda_{\rm UV}}{\tilde m } 
 =3 \times 10^{-5}  \left(\frac{2~{\rm TeV} }{\tilde m} \right)  \left( \frac{\log\Lambda_{\rm UV}/\tilde m}{30}\right) 
 \left(\frac{\tan\beta}{10}\right)^2~,
\ee
where we have omitted ${\cal O}(1)$ loop functions depending on mass ratios of heavy squarks and charginos, 
whose average mass  is denoted by $\tilde m$.

The $\tilde{t}_1 \to N b\ell^+\nu_\ell$ decay receives 
contributions suppressed by heavy sparticles or mediated only by virtual SM particles.  We here focus on
the latter contribution, which is dominant in our case. This leads to \eq{eq:4body}, whose derivation 
is given in the appendix, together with the
matrix element relevant for implementation in Monte Carlo codes.

The two decay channels can dominate in different regions of the parameter space and become roughly comparable for $\theta_{tc}\sim 10^{-5}$ and stop--neutralino mass differences motivated by DM considerations. However, the large model dependence of $\theta_{tc}$ prevents us from making any firm conclusion. The four-body decay has a much steeper dependence on $\Delta M$ and becomes less relevant for very small $\Delta M$. The decay $\tilde{t}_1 \to N b\ell^+\nu_\ell$, not previously considered in the literature, is interesting from the experimental point of view since it leads to an additional soft lepton.

Since the $\tilde{t}_1\to c N$ decay only produces an unobservable soft jet and its signatures have been previously studied,
in the following we focus our attention mainly on the four-body decay channels in eqs.~(\ref{eq:4body})--(\ref{eq:4body2}), 
which we assume to be  the dominant decay modes. 
As we discuss below, the bounds we are able to derive at present from existing LHC searches
are largely independent from this assumption; however, the presence of a lepton in the final state 
could possibly lead to stringent bounds with future optimized searches (see sect.~\ref{sect:dedicated}).
Assuming the four-body modes to be dominant
implies 
\beq 
\hbox{BR}(\tilde{t}_1 \to N b\ell^+\nu_\ell) \approx  1/9
\eeq
for each lepton flavor $\ell$. Moreover, the smallness of the total decay width 
implies that the decay vertex displacement may be detectable.


%

\subsection{Bounds from existing LHC searches}

The challenge of detecting stop decays for compressed spectra is all
in the capability of reconstructing and identifying the soft decay
products of the two stop decays. On the other hand, this is not the
only experimental handle we have. 

The first problem is with triggering these events. The jets and
leptons originating from the stop decay are too soft to be used to
retain the events during on-line selection, given the (CPU and bandwidth) 
budget of the experiments. The only possibility is to detect these
processes through the associate jet production ($\tilde t \tilde t^*$ plus one or more jets), with a consequent reduction of the effective cross section. 

In the worst case scenario, all the decay products are lost and one is
left with one or more jets:
bounds exist from monojet searches~\cite{monojetCMS,monojetATLAS} 
(performed in the Dark Matter context) and 
from searches with $\geq 2$ jets~\cite{ATLAS0lep,CMSalphaT} (performed in supersymmetric contexts);
see also~\cite{ATLASlj}. 

In the best case scenario one can also detect the leptons from the
decay of the stop pair. This is why it is interesting to consider a
set of analyses that focus on $\geq 2$ jets
for events with or without
leptons. The CMS razor analysis~\cite{razorVar,cmsrazor} is a
all-in-one answer to our needs, with the additional advantage that the
jet selection in the analysis is looser than the one used in the hadronic
SUSY searches: $p_T^{\rm jet}>60$ GeV for the first two jet;
$p_T^{\rm jet}>40$ for the other jets.  The looser jet selection increases
the effective cross section we are sensitive to.

To estimate the sensitivity of the search to the soft leptons from
the stop decays, we implemented an emulation of the razor analysis,
based on generator-level jets and leptons. We generate pair-produced
stop squarks in $\sqrt{s} = 7$ TeV pp collisions using {\tt
  PYTHIA8}~\cite{pythia8}. The stop are forced to decay with a flat
matrix element as $\tilde t \to \ell \nu_\ell b N$.
The transverse momenta of all the visible particles are summed to
compute the missing transverse energy at generator level. Similarly,
these particles are clustered into jets using the {\tt
  FASTJET}~\cite{FASTJET1,FASTJET2} implementation of the
antikT~\cite{antikt} algorithm. As for CMS, we use $R=0.5$ to define
the jet size. The razor variables and the six {\it boxes} (MuEle,
MuMu, EleEle, Mu, Ele, and Had) are defined following the instructions
provided by the CMS collaboration~\cite{razorLikelihood}.
To take into account the limited efficiency in lepton detection, we
applied the efficiency curves of the CMS dilepton SUSY
search~\cite{CMSdilepton}, using a hit-or-miss analysis. This is a
valid procedure, since the lepton definition in the razor and dilepton
SUSY searches are similar.

\begin{figure}[t]
\begin{center}
\includegraphics[width=0.32\textwidth]{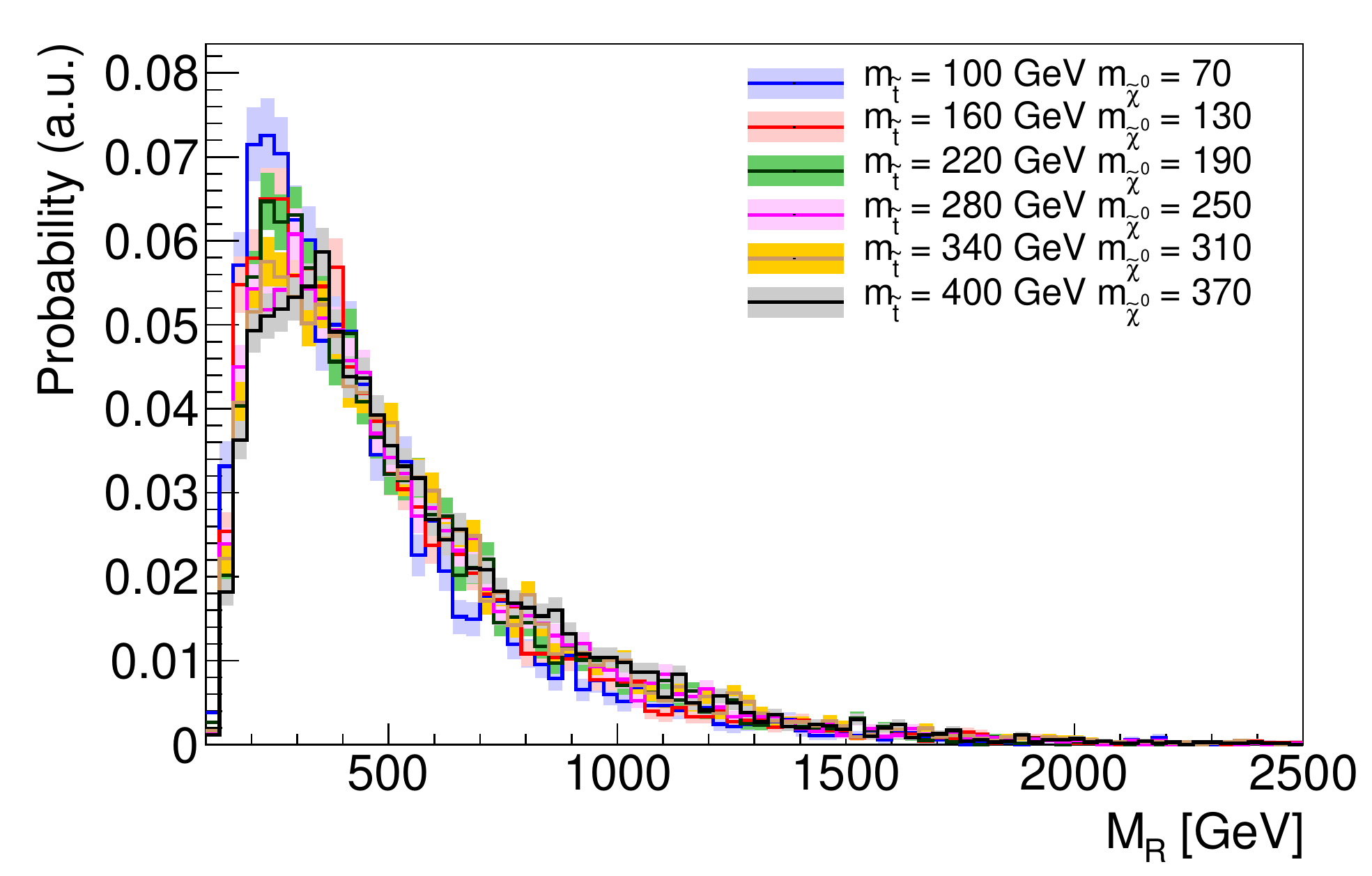}
\includegraphics[width=0.32\textwidth]{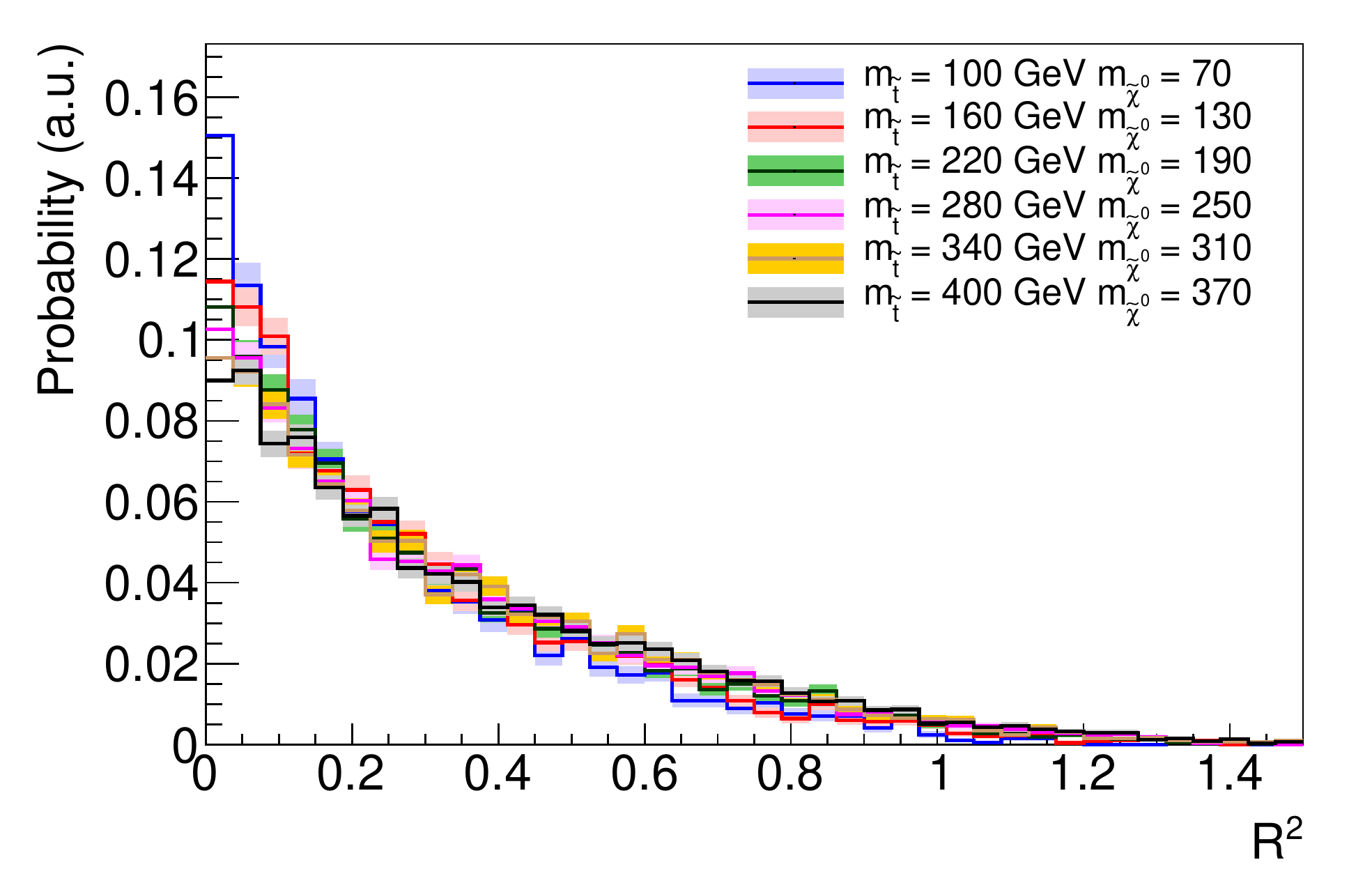}
\includegraphics[width=0.32\textwidth]{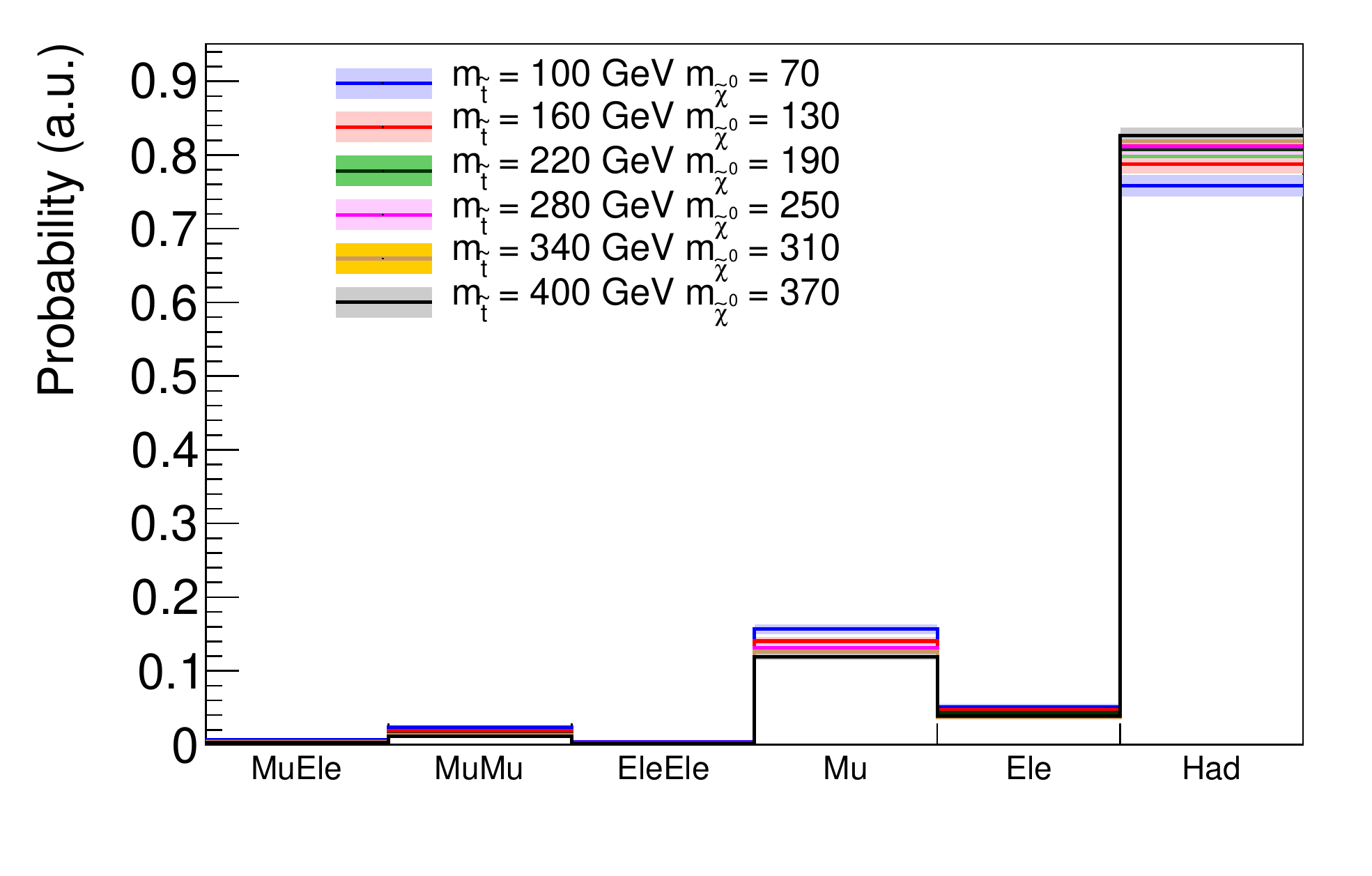}
\end{center}
\caption{\label{fig:razorVars}\em Distribution of $M_R$ (left), $R^2$
  (center), and box-by-box event fraction (right) for pair-produced
  stop events as a function of the stop mass, for $\tilde t \to \ell
  \nu_\ell b N$ decays and $m_{\tilde t} -   M_{\rm DM}= 30$ GeV. Even if this case is the most favorable for the
  selection of leptonic final states, the hadronic box is the most
  populated due to the small value of $m_{\tilde t} -   M_{\rm DM}$.}
\end{figure}

We scan the value of the stop mass between 100 GeV and 400 GeV, fixing
the stop-to-neutralino mass gap to 30 GeV. We show in
fig.~\ref{fig:razorVars} the distribution of the razor variables for
different stop masses, as well as the breakdown in boxes.  A few
important features should be noticed: 
\begin{itemize}
\item[i)] The $M_R$ variable approximates the momentum of the jets in
  the frame such that $|p_{j_1}|=|p_{j_2}|$. In the case of squark
  pair-production, for which this variable was designed, this
  corresponds to the squarks rest frame. This is why the $M_R$
  distribution for this case peaks at the $M_\Delta =
  (\mstopl^2-M_{\rm DM}^2)/{\mstopl}$. Instead, in the case we consider
  here the jets come from the associated (non-resonant) production and
  the peak is at $\sim 150$ GeV, regardless of the stop and neutralino
  masses (due to the selection on the jet $p_T$ and not to the SUSY
  kinematic).
\item[ii)] The $R$ variable is defined as $M_R^T/M_R$ where $M^T_R\le
  M_\Delta$ is a transverse invariant mass, such that the QCD
  background peaks at $R \sim 0$, while the signal can produce events
  with larger values of $R$, where the two jets have similar
  directions, opposite to the direction of the neutralinos. In the case
  of compressed stop spectra the $R^2$ distribution has some
  dependence on the stop mass, due to the correlation between the stop
  mass (setting the scale of the hard interaction), the spectrum of
  the associated jets, and the missing energy in the event.
\item[iii)] The majority
of the events selected by the analysis falls in the hadronic box. 
\end{itemize}
All
these features are explained by the fact that the analysis is only
sensitive to the events with two associated jets. These jets form the
hemispheres and the razor variables are computed for a {\it
  non-resonant} production. In the largest fraction of the events the
decay products of the stop are not seen and, effectively, the signal
behaves like for the direct production of Dark Matter~\cite{RazorDM},
The stop plays the part of the Dark Matter, with the big advantage of
the production cross section much larger than for Dark Matter direct
detection. At the same time, the result is largely independent on the
final state the stop decays to.

\begin{figure}[t]
\begin{center}
\includegraphics[width=0.45\textwidth]{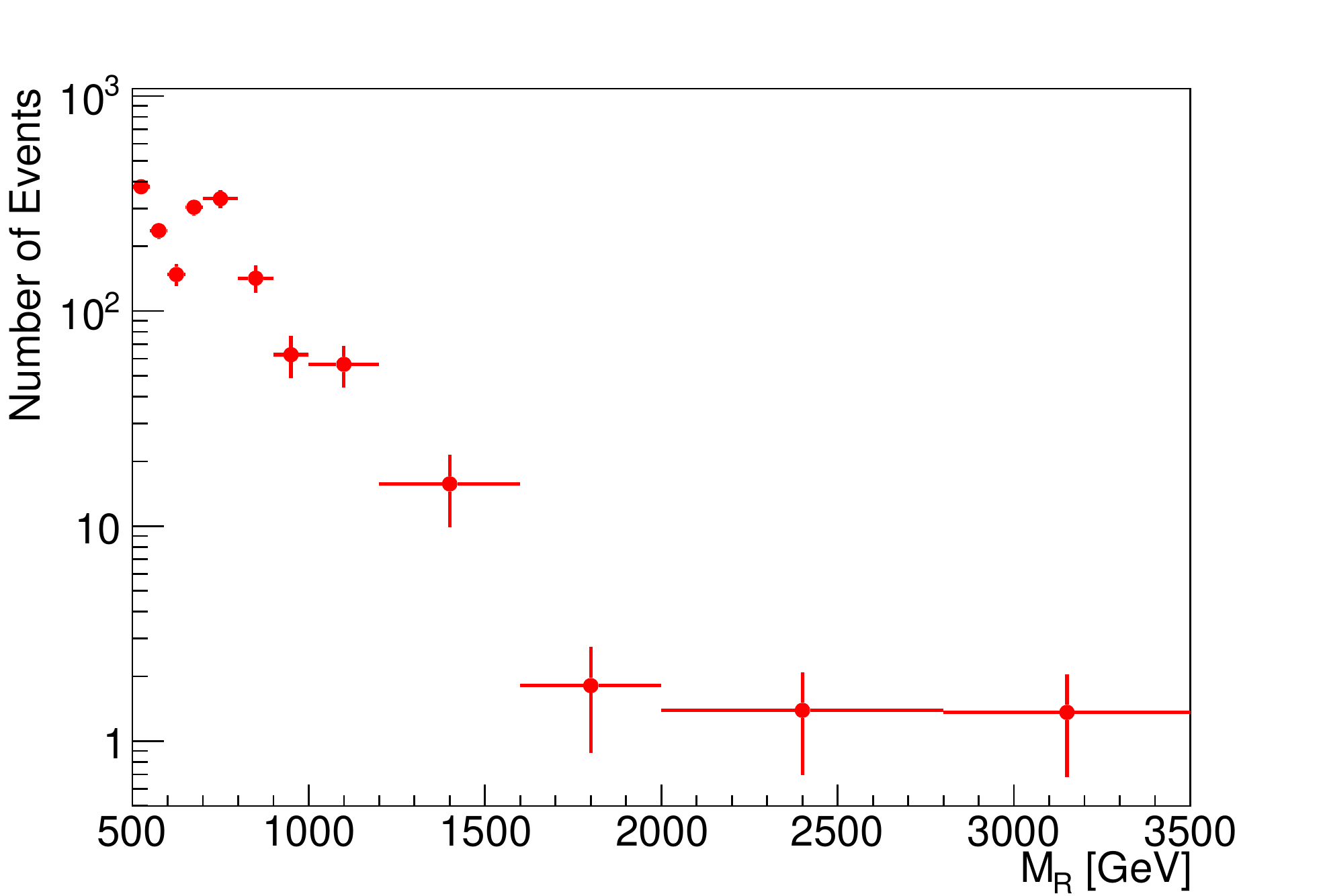}
\includegraphics[width=0.45\textwidth]{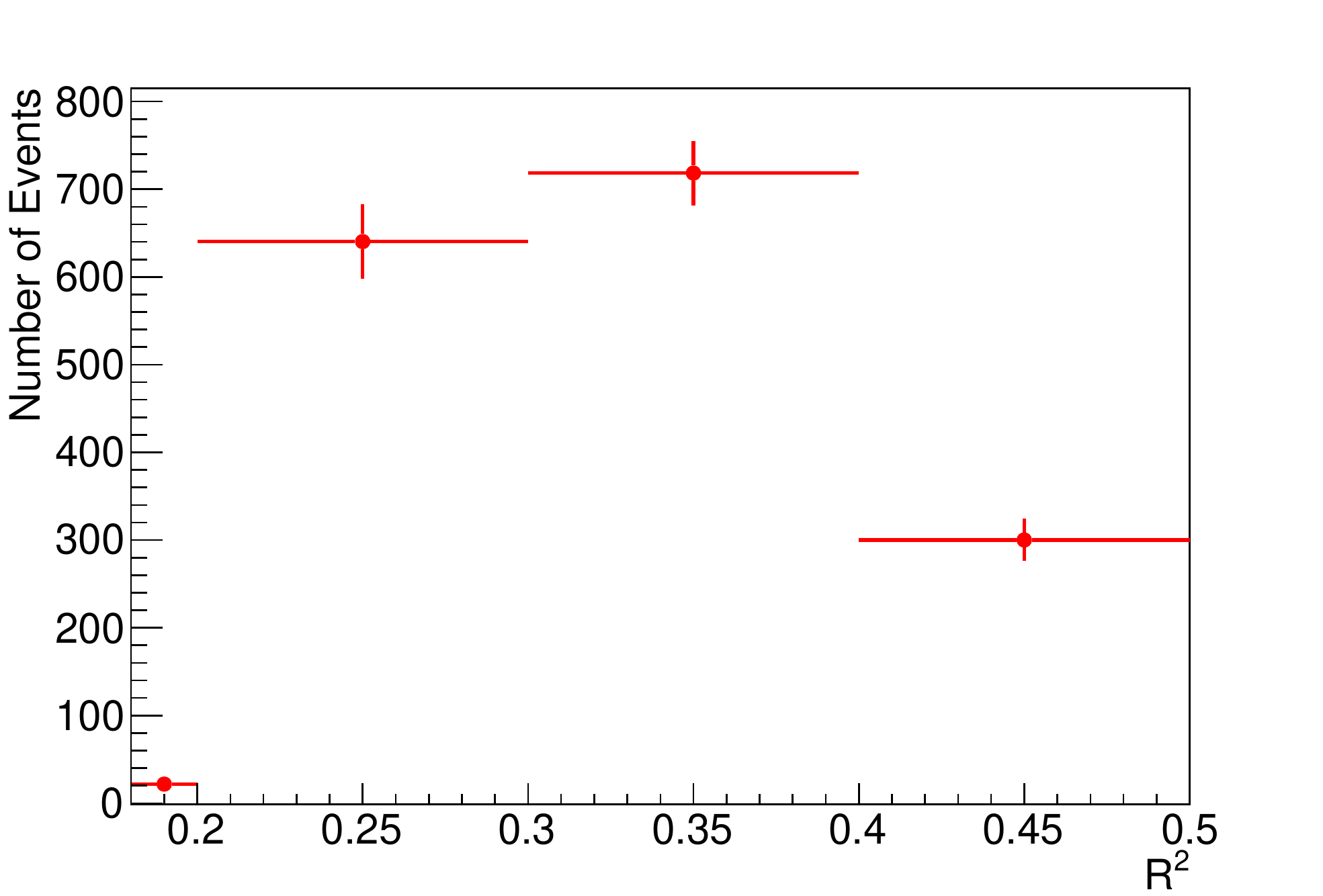}
\end{center}
\caption{\label{fig:HadB_bkg}\em
Projections of the expected background in the razor hadronic box, obtained from the bin-by-bin expected background
in the $\sqrt{s}=7$~TeV run of CMS (from ref.~\cite{cmsrazor,razorLikelihood}).}
\end{figure}

These considerations suggest that the Had box is the only relevant
sample to consider in our study. This is also the only
box for which the information needed for phenomenology studies
(observed yield and expected background vs $R^2$ and $M_R$) are
given (the number of expected background events is shown in fig.~\ref{fig:HadB_bkg}).
While we limit our study to the Had box, we stress the fact
that there is some sensitivity in the Mu box and Ele boxes, which 
could have be exploited if we had the relevant information.
The importance of the Had box over the others also implies that the
monojet analysis is a good candidate to look for our signal, as it is
for Dark Matter direct production. The

\begin{figure}[t]
\begin{center}
\includegraphics[width=0.48\textwidth,height=0.48\textwidth]{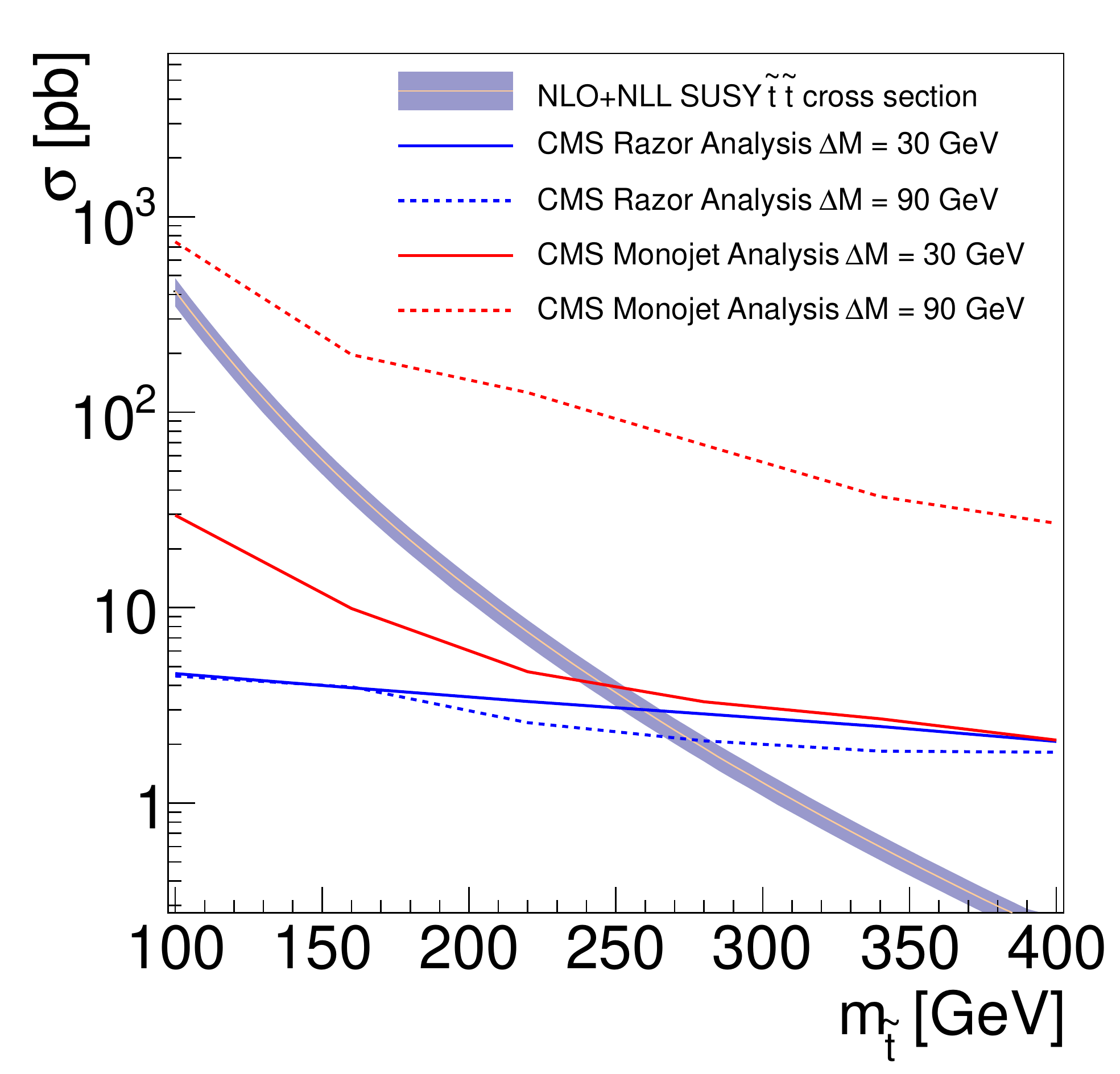}
\includegraphics[width=0.48\textwidth]{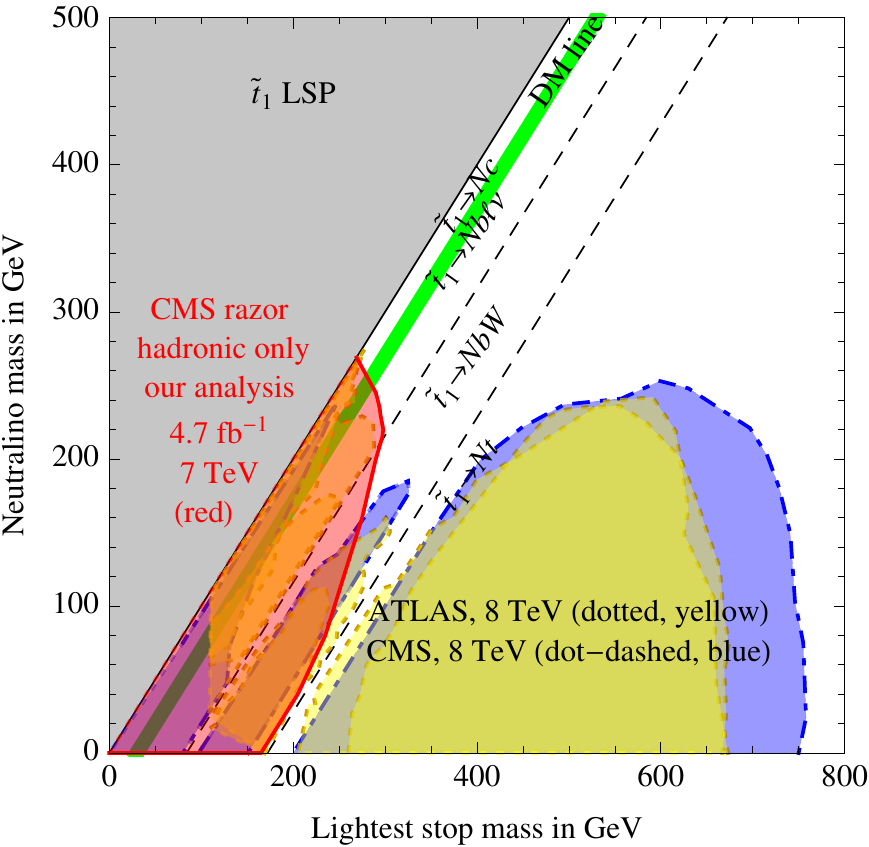}
\end{center}
\caption{\label{fig:limits}\em
{\bf Left}: predicted cross section and experimental limits as functions of the lightest stop mass.
{\bf Right}: excluded regions in the ($\mstopl,  M_{\rm DM}$) plane from 
our re-analysis of $7\TeV$ data (red) compared with
latest ATLAS  (dotted regions shaded  in yellow) and CMS  (dot-dashed regions shaded in green)
analyses of $8\TeV$ data.}
\end{figure}

We show in fig.~\ref{fig:limits} the limits obtained with the monojet
and the razor (Had-only) analyses. For both the analyses, we consider
the expected background yield (with error) and the observed yield, and
we model the likelihood according to a Poisson distribution. The
background uncertainty is described using a log-normal function. We
assign a $30\%$ error to the signal efficiency, to take into account
the differences between our implementation of the analysis and a more
realistic description of the CMS detector. We then derive a
posterior-probability density function for the signal cross section
as:
\begin{equation}
P(\sigma) = \int_0^\infty db \int_0^1 d\epsilon \frac{(b+L \sigma
  \epsilon)^n e^{-b-L \sigma \epsilon}}{n!} \hbox{Ln}(\epsilon| \bar
\epsilon, \delta_\epsilon)\,  \hbox{Ln}(b | \bar b, \delta_b)
\end{equation}
where $b$ ($\epsilon$) is the actual value for the background yield
(the efficiency), $\bar b$ ($\bar \epsilon$) is its expected value,
and $ \delta_b$ ($ \delta_\epsilon$) the associated error;
Ln$(x|m,\delta)$ is a log-normal function for $x$ with mean $m$ and
variance $\sigma$; $n$ is the observed yield, $L$ is the available
luminosity (for which we neglect the $\sim 4\%$ error) and $\sigma$ is
the signal cross section. In the case of the razor analysis, the
actual posterior is obtained as the product of the posteriors in each
of the bins provided in~\cite{razorLikelihood}. We verified that
taking $L=\epsilon=1$ we can reproduce the limit on the signal
strength for the two analyses. The 95\%-probability limit is obtained integrating 
the posterior from 0 up to the value $\sigma_{\rm UP}$ such that
\begin{equation}
\frac{\int_0^{\sigma_{\rm UP}} P(\sigma) d\sigma}{\int_0^\infty
P(\sigma) d\sigma} = 0.95.
\end{equation}

The left plot of fig.~\ref{fig:limits} shows the 95\%-probability
limit on the signal cross section as a function of the stop mass for
both the analyses, fixing the mass split at $30 \GeV$ and $100 \GeV$
(with the stop decayed to $t^* N$). The sensitivity of the
monojet analysis is limited by the tight selection on jets and missing
transverse energy. The limit is worse for larger splitting because of
the veto on any third jet with $p_T>30 \GeV$. At the contrary, the
razor analysis is more efficient for this signature and more
performant for larger splitting, since no veto is applied. One should
also consider that at large values of the mass splitting the five leptonic
boxes could further improve the sensitivity. 

The right plot of fig.~\ref{fig:limits} shows the limit in the stop mass vs neutralino mass
plane. This plot shows the same qualitative
features as the 1D limit plot. At large splits, the limit from the
razor analysis is found to be consistent with (and slightly worse
than) the official limit on stop pair production~\cite{cmsrazor}.
Both the 1D and 2D limits were obtained comparing the excluded cross
section with the NLO+NLL $\tilde t \tilde t^*$ cross section at 7 TeV
taking the decoupling limit for the other SUSY
particles~\cite{NLONNLxsec}.\footnote{In the revised version of the plot
(september 2014) we subtract the signal contribution in the sideband
to the background estimate by CMS. This effect, generically negligible in the models considered
by the original analysis, becomes relevant in our study for large values of the stop-neutralino mass splitting.
Furthermore, we plot the latest bounds from ATLAS and CMS with 8 TeV data.}

\subsection{Dedicated analyses}
\label{sect:dedicated}

The existing limit is interesting, considering how challenging this
signature is. This study also shows once more that the inclusive
searches by ATLAS and CMS are much more general than the signal
signatures they have been designed for. While a dedicated search could
do better for a specific scenario, the inclusive searches are a good
assurance policy for unexpected signatures. Repeating the analysis at
8 TeV with more data will certainly push the sensitivity further. On
the other hand, we think it is interesting to imagine how the analyses
could be changed to improve the sensitivity.

One could certainly gain by using looser kinematic requirements. The
limiting factor is related to the triggers. For example, it was
pointed out extending the razor analysis at the tail of $R^2$ for low
$M_R$ could improve the sensitivity to DM
production~\cite{RazorDM}. The same conclusion applies to compressed
stop-neutralino spectra, since the signature in the razor Had box is
the same. 

\begin{figure}[t]
\begin{center}
\includegraphics[width=0.48\textwidth,height=0.48\textwidth]{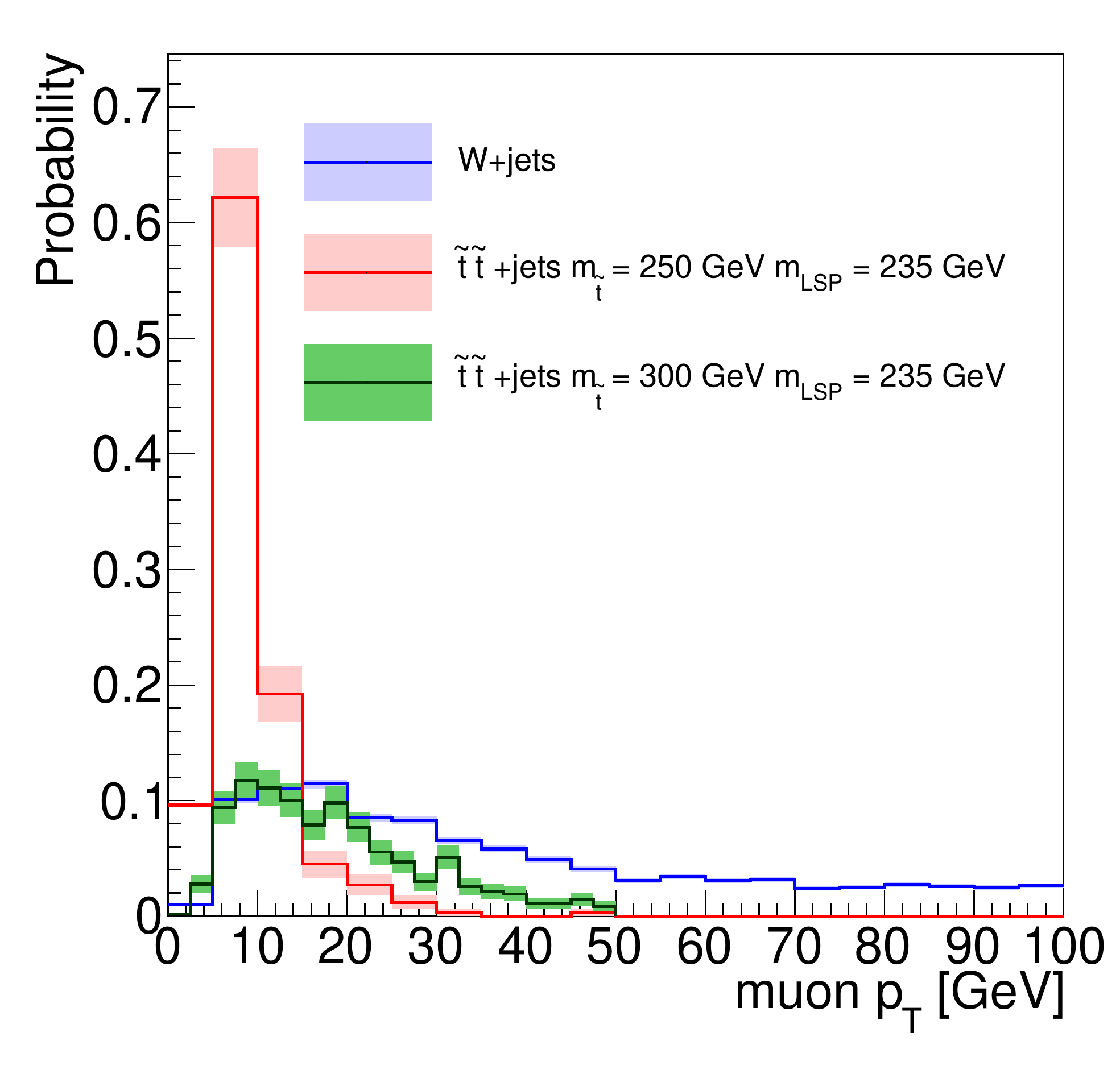}
\includegraphics[width=0.48\textwidth,height=0.48\textwidth]{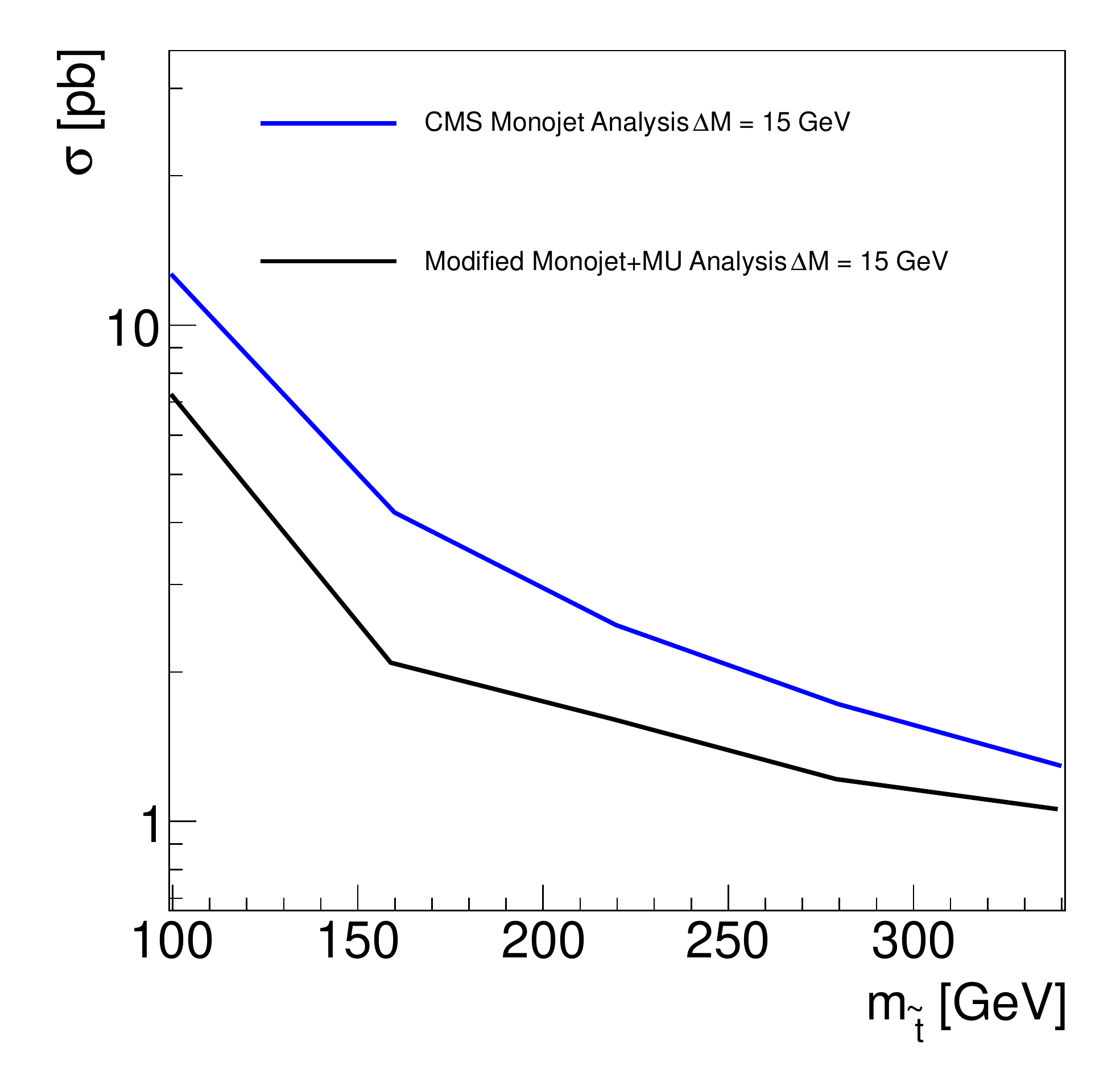}
\end{center}
\caption{\label{fig:monojetModified}\em Improvements that can be obtained with a dedicated search.
{\bf Left}: distribution of
  the muon $p_T$ for $W$+jets and stop-pair events passing the CMS
  monojet selection criteria, except for the muon veto and the veto on
  isolated tracks.  {\bf Right}: 
  expected excluded cross section for stop pair
  production obtained from the CMS monojet analysis (blue) and a
  modified monojet+muon search (black). Events are generated with
  four-body stop decays to $f \bar f' b N$, of which $\sim 20\%$
  produce one muon.}
\end{figure}

A change in the lepton selection could further increase the
sensitivity of these analyses. The left plot on
Fig.~\ref{fig:monojetModified} shows the distribution of the muon
$p_T$ for $W$+jets events selected by the CMS monojet analysis, before
applying the muon veto and the isolated track veto. This is compared
to the equivalent distribution obtained for events with pair-produced
stops, decaying to $W^* b  N$, with at least one of the two $W^*$
producing a $\mu \nu$ pair. We consider two values of the stop mass
($m_{\tilde t} = 150$ GeV and $m_{\tilde t} = 270$ GeV) for $\Delta M
= 15$ GeV. Requiring one muon with $p_T<15$ GeV corresponds to
reducing the $Z(\nu\nu)$+jets background to a negligible level, and to
rejecting $\sim 92\%$ of the other backgrounds. 

To evaluate the
potential improvement due to this change, we applied the monojet
analysis to the generated stop-stop samples, and we separate the
selected events in two boxes (as for the razor analysis): the Muon
box, including all the events with one muon with $p_T<15$ GeV; the Had
box, with all the other events. We then distribute the background in
the two boxes as follows: all the $Z(\nu\nu)$+jets background to the
Had box; $8\%$ ($92\%$) of the other background in the Mu (Had) box.
We then evaluate the potential sensitivity of this modified analysis
on a sample of pair-produced stop decays, decaying to $W^* b N$,
$20\%$ of which produce at least one muon in stop decay. 

The right
plot of Fig.~\ref{fig:monojetModified} shows the expected exclusion
limit, compared to what is obtained with the usual monojet analysis. A
similar improvement could be used for electrons, provided the
understanding of the electron identification and the fake rate at low
$p_T$.  One should keep in mind that our results come from a
simplified description of the CMS detector. A more accurate
assessment of the improvement can only be obtained with a detailed
simulation of the detector performances. We look forward to see this
change applied to the monojet analyses by ATLAS and CMS.

As a side remark, we would like to stress the fact that the stop decay
products could be displaced from the primary vertex of the
proton-proton collision. Requiring a displaced vertex, particularly
with one muon originating from it, can potentially reduce the standard
model background to a very small level. However for $\Delta M\approx
30\GeV$ only a small fraction of the $\tilde{t}_1$ decay after a
detectable path of about 1 mm. An accurate estimate of the signal
sensitivity for a diplaced-vertex analysis would require an accurate
description of the vertex resolution for the LHC detectors and should
be investigated directly by the ATLAS and CMS collaborations. Even if
the signal reduction is too large to be beneficial for 8 TeV searches,
this is an interesting possibility in light of the high statistics expected
for the future 14 TeV LHC run.


\section{Conclusions}

We have put forward a series of theoretical arguments that motivate the existence of a mainly right-handed stop in the mass range $\mstopl = 200$--$400\gev$, together with a neutralino 30--$40\gev$ lighter and, possibly, a gluino with mass below 1.5~TeV. However, quite independently of any of these specific motivations, the search for stops nearly degenerate with the neutralino LSP is an important experimental task, necessary to cover possible corners of parameter space where supersymmetry may still hide.

We have pointed out that, when the mass splitting between stop and neutralino is smaller than $M_W+m_b$, the previously-neglected four-body decay processes $\tilde{t}_1 \to N b\ell^+\nu_\ell$ and $\tilde{t}_1 \to N b q \bar q^\prime$ can compete with the flavor-changing decay $\tilde{t}_1 \to cN$. The presence of a charged lepton in the final state of the four-body decay gives a useful handle to identify the stop in this experimentally difficult mass configuration where all visible particles are relatively soft.

Regardless of the particular stop decay mode, the request of an extra jet in association with the stop pair greatly improves triggering capability and signal identification. We have shown that the inclusive searches using razor analysis are very efficient to probe stops nearly degenerate with neutralinos. In this region of mass parameters, we are able to set limits on the stop that are stronger than those published by ATLAS and CMS. Our limits (shown in fig.~\ref{fig:limits}) extend up to stop masses of about 250~GeV, even for a vanishing stop--neutralino mass difference. This means that the LHC has already started to probe the ``light stop window" motivated by our theoretical considerations, but most of the interesting region will be explored only at LHC14.

\small

\subsubsection*{Note Added}
While our paper was being completed, similar results were presented in~\cite{cina,usa}. 
In our analysis we  used the
response function for the CMS detector, provided by the CMS
collaboration, instead of trying to emulate the LHC detectors.
More importantly, we used  the full likelihood provided by the CMS
collaboration for the inclusive razor analysis, which gives a
realistic description of the likelihood resulting from  the data.  This
prevented us from extending our study to the razor btag
search~\cite{cmsrazorb}. The latter has a better sensitivity in
presence of bjets with $p_T>40$ GeV (i.e. far from the diagonal of the
$m_{\tilde \chi^0}$ vs. $m_{\tilde t}$ plane), if an accurate
emulation of the btag efficiency and the mistag rate is reached. On the diagonal,
only a small fraction of the associated jets are bjets, such that
requiring a btag has the effect of reducing the expected signal much
more than the factor-three background reduction. 

\subsubsection*{Acknowledgments} AD was partly supported by the National Science Foundation
under grants PHY-0905383-ARRA and PHY-1215979. 
AS was supported by the ESF grant MTT8 and by
SF0690030s09 project.
 
\section*{Appendix}
In this appendix we derive the four-body stop decay width given in \eq{eq:4body}.
In the limit of small mass difference ($\Delta M\ll M_W,m_t$), the
amplitude for pure right-handed stop decay $\tilde{t}_R \to N b
e^+\nu$ is 
\be
|\mathscr{A}|^2= \frac{32 g^6 \tan^2 \theta_{\rm W}}{9M_W^4 m_t^2} (P_N \cdot
P_e)(P_b\cdot P_\nu) ~,
\ee
where $P_i$ are the quadri-momenta of
the particles involved.  The decay width is given by 
\be
\Gamma = \int
d\phi^{(4)} \frac{|\mathscr{A}|^2}{2 m_{\tilde t}} ~.
\ee  
The 4-body phase space
integral $d\phi^{(4)}$ can be analytically performed at leading order
in $\Delta M =m_{\tilde{t}}- M_{\rm DM}$.  Indeed, by writing the decay as $\tilde{t}\to
XY\to(Ne)(b\nu)$, the amplitude for each sub-decay is separately Lorentz
invariant. Thus, using 
\beq 
d\phi^{(4)} = \frac{ds_X ds_Y}
{(2\pi)^2} 
d\phi^{(2)}(\tilde{t}\to XY) 
d\phi^{(2)}(X\to Ne) d\phi^{(2)}(Y\to b\nu) ~,
\eeq  
\beq 
|\mathscr{A}|^2
=\frac{8  g^6  \tan^2 \theta_{\rm W}}{9M_W^4 m_t^2} \,{s_X}\, ({s_Y -  M_{\rm DM}^2}) ~, 
\eeq 
we
get
\be
\Gamma =  \int_{ M_{\rm DM}^2}^{m_{\tilde{t}}^2} ds_Y  
 \int_0^{(m_{\tilde{t}}-\sqrt{s_Y})^2} ds_X
 \left(1-\frac{ M_{\rm DM}^2}{s_Y}\right)\frac{
 \lambda(m_{\tilde t}^2,s_X,s_Y) |\mathscr{A}|^2}{4(4\pi)^5m_{\tilde t}^3}
=\frac{2g^6  \tan^2 \theta_{\rm W}\, I}{9(4\pi)^5M_W^4m_t^2 m_{\tilde t}} ~,
\ee
\be
I\equiv m_{\tilde t}^8\int_{ M_{\rm DM}^2/m_{\tilde t}^2}^1 dy\int_0^{(1-\sqrt{y})^2}dx~\frac xy \left( y-\frac{ M_{\rm DM}^2}{m_{\tilde t}^2}\right)^2\sqrt{(1+x-y)^2-4x}\approx \frac{8\, \Delta M^8}{315} ~,
\ee
where we have kept only the leading order in $\Delta M $. From these expressions we obtain \eq{eq:4body}.

%
%
%


\begin{thebibliography}{99}
{\footnotesize


\bibitem{HiggsATLAS}
ATLAS Collaboration,  Phys.\ Lett.\ B {716} (2012) 1 [arXiv:1207.7214].

  
  \bibitem{HiggsCMS}
 CMS Collaboration,  Phys.\ Lett.\ B {716} (2012) 30  [arXiv:1207.7235].

\bibitem{nat}
See e.g.\
A.~Strumia,
  JHEP {1104} (2011) 073
  [arXiv:\hhref{1101.2195}].



\bibitem{Barbieri:2009ev}
  R.~Barbieri and D.~Pappadopulo,
  JHEP {0910} (2009) 061
  [arXiv:0906.4546].

\bibitem{Papucci:2011wy}
  M.~Papucci, J.~T.~Ruderman and A.~Weiler,
  JHEP {1209} (2012) 035
  [arXiv:1110.6926].

\bibitem{Brust:2011tb}
  C.~Brust, A.~Katz, S.~Lawrence and R.~Sundrum,
  JHEP {1203} (2012) 103
  [arXiv:1110.6670].
  
\bibitem{Han:2012fw}
  Z.~Han, A.~Katz, D.~Krohn and M.~Reece,
  JHEP {1208} (2012) 083
  [arXiv:1205.5808].

\bibitem{Espinosa:2012in}
  J.~R.~Espinosa, C.~Grojean, V.~Sanz and M.~Trott,
  arXiv:1207.7355 [hep-ph].



\bibitem{Carena:2008rt}
 M.~Carena, G.~Nardini, M.~Quiros and C.~E.~M.~Wagner,
 JHEP {\bf 0810} (2008) 062
 [arXiv:0806.4297].
 
 
\bibitem{Exppages}
https://twiki.cern.ch/twiki/bin/view/AtlasPublic\\
http://cms.web.cern.ch/org/cms-papers-and-results




\bibitem{Chou:1999zb}
  C.~-L.~Chou and M.~E.~Peskin,
  Phys.\ Rev.\ D {61} (2000) 055004
  [hep-ph/9909536].
  

\bibitem{Hiller:2009ii}
  G.~Hiller, J.~S.~Kim and H.~Sedello,
  Phys.\ Rev.\ D {\bf 80} (2009) 115016
  [arXiv:0910.2124 [hep-ph]].

  
\bibitem{Alves:2012ft}
  D.~S.~M.~Alves, M.~R.~Buckley, P.~J.~Fox, J.~D.~Lykken and C.~-T.~Yu,
  arXiv:1205.5805.

\bibitem{Kilic:2012kw}
  C.~Kilic and B.~Tweedie,
  arXiv:1211.6106.



  
  
    
  \bibitem{fit}
  The result is obtained updating the fit of
  P.~P.~Giardino, K.~Kannike, M.~Raidal and A.~Strumia,
  arXiv:1207.1347.
  
  
\bibitem{Ciuchini:1998xy}
  M.~Ciuchini, G.~Degrassi, P.~Gambino and G.~F.~Giudice,
  Nucl.\ Phys.\ B {534} (1998) 3
  [hep-ph/9806308].
  
\bibitem{ASold}
A.~Strumia,
  Phys.\ Lett.\ B {397} (1997) 204
  [\hhref{hep-ph/9609286}].
  
  
\bibitem{Gabrielli:1994ff}
  E.~Gabrielli and G.~F.~Giudice,
  Nucl.\ Phys.\ B {433} (1995) 3
   [Erratum-ibid.\ B {507} (1997) 549]
  [hep-lat/9407029].


\bibitem{Misiak:2006zs}
  M.~Misiak, H.~M.~Asatrian, K.~Bieri, M.~Czakon, A.~Czarnecki, T.~Ewerth, A.~Ferroglia and P.~Gambino {\it et al.},
  Phys.\ Rev.\ Lett.\  {98} (2007) 022002
  [hep-ph/0609232].

\bibitem{hfag}
Heavy Flavor Averaging Group Collaboration,
  arXiv:1207.1158.
  

\bibitem{Babar}
BABAR Collaboration,
  arXiv:1207.5772.
  
    
\bibitem{UTfit}
  A.~J.~Bevan {\it et al.},
  PoS HQL {2010} (2011) 019 [http://utfit.org/UTfit].


  
  
  
  \bibitem{DM}  C.~Boehm, A.~Djouadi and Y.~Mambrini,
  Phys.\ Rev.\ D {61} (2000) 095006
  [hep-ph/9907428].
        
       \bibitem{Som}   A.~De Simone, G.~F.~Giudice and A.~Strumia,
  JHEP {1406} (2014) 081
  [arXiv:\hhref{1402.6287}].
        
     \bibitem{CMSSMscan}
     M.~Farina, M.~Kadastik, D.~Pappadopulo, J.~Pata, M.~Raidal and A.~Strumia,
  Nucl.\ Phys.\ B {853} (2011) 607
  [arXiv:1104.3572].   
  M.~Kadastik, K.~Kannike, A.~Racioppi and M.~Raidal,
  JHEP {1205} (2012) 061
  [arXiv:1112.3647].
        
\bibitem{Hiller:2008wp}
  G.~Hiller and Y.~Nir,
  JHEP {0803} (2008) 046
  [arXiv:0802.0916].


\bibitem{monojetCMS}
CMS Collaboration,
  JHEP {1209}, 094 (2012)
  [arXiv:1206.5663].

\bibitem{monojetATLAS}
ATLAS Collaboration,
  arXiv:1210.4491.

\bibitem{ATLAS0lep}
ATLAS Collaboration,
  Phys.\ Lett.\ B {710}, 67 (2012)
  [arXiv:1109.6572].
  
  \bibitem{ATLASlj}
  \href{http://cdsweb.cern.ch/record/1497732/files/ATLAS-CONF-2012-166.pdf}{ATLAS conference note 2012-166}.
  
  

\bibitem{CMSalphaT}
CMS Collaboration,
  arXiv:1210.8115.

\bibitem{razorVar}
 C.~Rogan,
  arXiv:1006.2727.

\bibitem{cmsrazor}
\hepart[1212.6961]{CMS Collaboration}.


\bibitem{pythia8}
T.~Sjostrand, S.~Mrenna and P.~Z.~Skands,
  Comput.\ Phys.\ Commun.\  {178}, 852 (2008)
  [arXiv:0710.3820].
  
\bibitem{FASTJET1}
M.~Cacciari, G.~P.~Salam and G.~Soyez,
  Eur.\ Phys.\ J.\ C {72}, 1896 (2012)
  [arXiv:1111.6097].

\bibitem{FASTJET2}
M.~Cacciari and G.~P.~Salam,
  Phys.\ Lett.\ B {641}, 57 (2006)
  [hep-ph/0512210].

\bibitem{antikt}
 M.~Cacciari, G.~P.~Salam and G.~Soyez,
  JHEP {0804}, 063 (2008)
  [arXiv:0802.1189].

\bibitem{razorLikelihood} The details on how to implement the CMS
  razor analyses outside the CMS analysis framework are given in {\tt
    https://twiki.cern.ch/twiki/bin/view/CMSPublic/RazorLikelihoodHowTo}.

\bibitem{CMSdilepton}
CMS Collaboration,
  JHEP {1208}, 110 (2012)
  [arXiv:1205.3933].

\bibitem{RazorDM}
 P.~J.~Fox, R.~Harnik, R.~Primulando and C.~-T.~Yu,
  Phys.\ Rev.\ D {86}, 015010 (2012)
  [arXiv:1203.1662].

\bibitem{NLONNLxsec}
 M.~Kramer, A.~Kulesza, R.~van der Leeuw, M.~Mangano, S.~Padhi, T.~Plehn and X.~Portell,
  arXiv:1206.2892.


\bibitem{cina}
\hepart[1211.2997]{Z.-H. Yu, X.J. Bi, Q.-S. Yan, P.-F. Yin}.

\bibitem{usa}
\hepart[1212.4856]{K. Kriza, A. Kumar, D.E. Morissey}.


\bibitem{cmsrazorb}
CMS Collaboration,
  CMS-PAS-SUS-11-024.


}
\end{thebibliography}
\end{document}